\documentstyle[11pt,paspconf,epsf]{article}
\newcommand{\rosat}{{\it ROSAT} }
\newcommand{\asca}{{\it ASCA} }

\newcommand{\ginga}{{\it GINGA} }

\newcommand{\pspc}{{\it PSPC} }
\newcommand{\etal}{{\it et al.} }
\newcommand{\bbxrt}{{\it BBXRT} }

\newcommand{\nh}{$\rm N_{H}$}

\begin{document}

\title{X-ray Observations of LINER and Starburst Galaxies}
\author{P. Serlemitsos, A. Ptak, and T. Yaqoob}
\affil{NASA GSFC/LHEA, Code 662, Greenbelt MD 20771}

\begin{abstract}
We present the results of \asca
observations of a heterogenous sample of
15 spiral galaxies. 8 
are LINERs or low-luminosity AGN (LLAGN), 5 are starburst galaxies and 2
are normal spiral galaxies.
We find that in all cases the \asca spectra can be described
by a canonical model consisting of a power-law with a photon index,
$\Gamma \sim 1.7-2.0$, plus a soft optically thin emission component
with $kT \sim 0.6-0.8$ keV. The implied element abundances are
often sub-solar.
The soft component is usually extended
and the nuclear, point-like emission is sometimes absorbed by
column densities in the range $\sim 10^{21} - 10^{23} \rm \ cm^{-2}$.
The relative luminosities of the soft and hard components vary from
galaxy to galaxy. For the LINERs,
the 2-10 keV luminosity of
the hard component is typically $\sim 10^{40-41} \rm \ ergs \ s^{-1}$
whereas the 0.5-2.0 keV
luminosity of the soft component is typically $\sim 10^{39-40} \rm \ ergs \
s^{-1}$. For starbursts, the 2-10 keV luminosity of
is $\sim 10^{39-40} \rm \ ergs \ s^{-1}$, somewhat lower than the 
corresponding luminosity of most of the LINERs in our sample.
The hard component is similar to the observed X-ray spectra of quasars
and also to the {\it intrinsic} X-ray spectra of 
classical Seyfert galaxies. Most of
the galaxies
in our sample exhibit no significant ($\Delta I/I > 20\%$) short-term 
variability (with timescales of a
day or less) whereas long-term variability is common.  We present a case
study of the LINER M81 in detail where there is evidence of large-amplitude
($\Delta I/I \sim 70\%$) variability over several weeks. There is also
clear evidence for a broad, complex Fe-K emission line which is
compatible with an origin in an accretion disk viewed at $\sim 40$
degrees. These results suggest a strong connection between
classical AGN, LINERs, and starburst galaxies.

\end{abstract}


\section{Introduction}
Optical emission lines are found in the nuclei of three types of galaxies: 
active galactic
nuclei (AGN), low-ionization emission line regions (LINERs; Heckman, 1980)
and starburst galaxies.  The dominant physical process
driving the emission
lines in AGN is thought to be accretion onto a supermassive (M $\sim \rm \ 10
^{5-9} M_{\sun}$) black-hole, while in starbursts the lines are the result of a
current or recent episode of star formation.
However, the physical origin of the line emission in LINERs is still unclear 
(see review by  A. Fillipenko in this volume).  
The two most likely scenarios are 
AGN-type accretion or starburst-type activity 
(i.e., shocks from supernovae).  X-ray 
observations provide important clues to distinguish 
between  them.
AGN are compact ($\ll 1$ pc), variable sources 
in which the X-ray emission has a nonthermal, power-law form.
In addition, Seyfert galaxies are characterized by strong
Fe K line emission produced by fluorescence in cold matter.
On the other hand, nuclear starbursts are frequently extended over kpc scales
and are expected to have
thermal spectra characterized by coronal X-ray emission lines.  

Below we shall see that the \asca (See Tanaka, Inoue
\& Holt 1994) spectra of low-luminosity
spiral galaxies are complex, usually requiring a two-component
description. Previous X-ray studies have been hampered by the lack of
instrument sensitivity, spectral resolution and restricted energy
bandpass. In particular, the {\it hard} X-ray spectra were studied
with non-imaging instruments where confusion of more than one bright
source in the galaxy was a hinderance.

The X-ray satellite
\asca, launched in 1993, is capable of 0.4-10 keV imaging, spectroscopy and
temporal analysis.   \asca is the first imaging X-ray satellite sensitive
above 4 keV and has better spectral resolution ($\rm \Delta E/E \sim 2\%$
at 6 keV) than any X-ray imaging mission to date. The
half-power radius of the X-ray mirrors is $\sim 1.'5$ and this is
sufficient to obtain individual
spectra of sources separated by $\sim 3-4'$ or more (typical of
the bright sources found in galaxies in our sample).
Here we present preliminary \asca results
on a sample of fifteen galaxies
(some previously published), eight of which are LINERS/LLAGN.  
When available,
some limited analysis was also performed with public \rosat data.

\section{Observations}

\begin{table}
\small
\begin{center}
\caption{Log of \asca and \rosat Observations}
\begin{tabular}{cccccccc}
\tableline 
Galaxy & Det.$^*$ & Date & Exp. & \nh$^\dagger$  
& Dist. & No. of Srcs. \\
& & & (ks) & & (Mpc) \\ %& $10^{-2} \rm \ cnts \ s^{-1}$ \\
\tableline
M33 & A & 7/22-23/93 & $\sim 40$ & $6.3^b$ & $1.2^a$ & 1 \\
& P & 1/7-9/93 & 16 & & & 1 \\ 
NGC 253 & A & 6/12-13/93 & 27-33 & $1.3^b$ & $2.5^h$ & 3 \\
& P & 12/25/91-6/05/92 & 23 & & & 5 \\
M81 & A & See \S 6 & & $4.3^b$ & $3.6^{j}$ & 3 \\
& P & 4/3-24/93 & 19 & & & 7 \\
M82 & A & 4/19-20/93 & 15-20 & $4.3^b$ & $3.6^{j}$ & 1 \\
& P & 3/28/91-10/16/91 & 23 & & & 1 \\
NGC 1313 & A & 7/13/93 & 28-29 & $3.7^g$ & $4.5^f$ & 3 \\
& P & 4/24/91-5/11/91 & 13 & & & 8 \\
NGC 6946 & A & 5/31-6/1/93  & 20-30 & $20-50^{b,i}$ & $8.3^d$ & 1 \\
& P & 6/16-21/92 & 37 & & & 9 \\
NGC 4258 & A & 5/15-16/93 & 36 & $1.2^b$ & $10.2^d$ & 1 \\
& P & 11/11-15/93 & 20 & & & 3 \\
M51 & A & 5/11-12/93 & 33-36 & $1.3^b$ & $14.0^a$ & 2 \\
& P & 11/28/91-12/13/91 & 22 & & & 10 \\
NGC 3628 & A & 12/12/93 & 18-20 & $2.0^c$ & $14.9^a$ & 2 \\ 
& P & 11/23-26/91 & 14 & & & 6 \\
NGC 3310 & A & 4/17/94 & 18-19 & $1.1^b$ & $19.4^a$ & 1 \\
& P & 11/17-18/91 & 9 & & & 5 \\
NGC 3998 & A & 5/10-11/94 & 32-39 & $1.2^a$ & $24.3^k$ & 1 \\
& P & 5/22-24/91 & 57 & & & 1 \\
NGC 4594 & A & 1/20/94 & 19-22 & $3.8^b$ & $22.6^a$ & 1 \\
& P & 7/15-19/92 & 11 & & & 3 \\
NGC 4579 & A & 6/25-26/95 & $\sim 40$ & $1.8^e$ & $36.1^a$ & 1 \\
& P & 12/15-16/91 & 9 & & & 1 \\
NGC 3079 & A & 5/09-10/93 & 29-38 & $0.8^b$ & $28.8^a$ & 1 \\
& P & 11/14-15/91 & 19 & & & 1 \\
NGC 3147 & A & 9/30/93 & 25-35 & $2.5^e$ & $56.4^a$ & 1 \\  
& P & 10/16-22/93 & 9 & & & 1 \\
\tableline
\tableline
\end{tabular}
\end{center}
\footnotesize
All distances assuming $H_0 \rm \ = \ 50 \ km \ s^{-1} \ Mpc^{-1}$ and 
$q_0 = 0$. \\ 
$^*$ A = \asca, P = \rosat \pspc \quad
$^\dagger$ Galactic column in units of $10^{20} \rm \ cm^{-2}$ \\
%$^\#$ See Ishisaki \etal 1996 \\
$^a$ Soifer, B. T., \etal 1987, ApJ, 320, 238 \\
$^b$ Fabbiano, G., \etal 1992, ApJS, 80, 531 \\
$^c$ Hartmann, D., \& Burton, W. 1995, In The Leiden-Dwingeloo Atlas of 
Galactic
Neutral Hydrogen (Cambridge: Cambridge University Press, UK), in press \\
$^d$ Tully, R. 1988, Nearby Galaxy Catalog (Cambridge University Press) \\
$^e$ Stark, A., \etal 1992, ApJS, 79, 77 \\
$^f$ de Vaucouleurs, G. 1963, ApJ, 137, 720 \\
$^g$ Cleary, M., \etal 1979, A\&AS, 36, 95 \\
$^h$ de Vaucouleurs, G. 1978, ApJ, 224, 710 and 
Davidge, T. \& Pritchett, C. 1990, AJ, 100, 102 \\
$^i$ Burnstein, D. \& Heiles, C. 1984, ApJS, 54, 33 \\
$^j$ Freedman, W., \etal 1994, ApJ, 427, 628 \\
$^k$ de Vaucouleurs, G., \etal 1991, Third Reference Catalogue of Bright 
Galaxies (New York: Springer-Verlag) \\
\normalsize
\end{table}
A log of the \asca and \rosat (\pspc only) observations is given in Table 1. 
The \asca data were reduced in a manner similar to the procedures described
for NGC 3147 (Ptak \etal 1996) and NGC 3628 (Yaqoob \etal 1995a).  Briefly,
\asca consists of two solid-state imaging spectrometers (SIS; hereafter
S0 and S1) and two gas 
imaging
spectrometers (GIS; hereafter G2 and G3).  
The spectra were typically accumulated from counts 
within 3' 
of the source centroid in the case of the SIS and 4' in the case of the 
GIS. In contrast to the SIS, the GIS introduces its own broadening to
the PSF of the X-ray mirrors. The
background was subtracted using counts in an annulus surrounding the source, 
typically
$\sim 6'-12'$.  Larger
source and background regions were used in the case of extended sources and for
some sources the spectra from different CCD chips were combined.  However, 
note that 
for most of the galaxies in this sample, statistical errors dominate
current instrumental uncertainties.

\section{Images}

\begin{figure}[htbp]
\plotfiddle{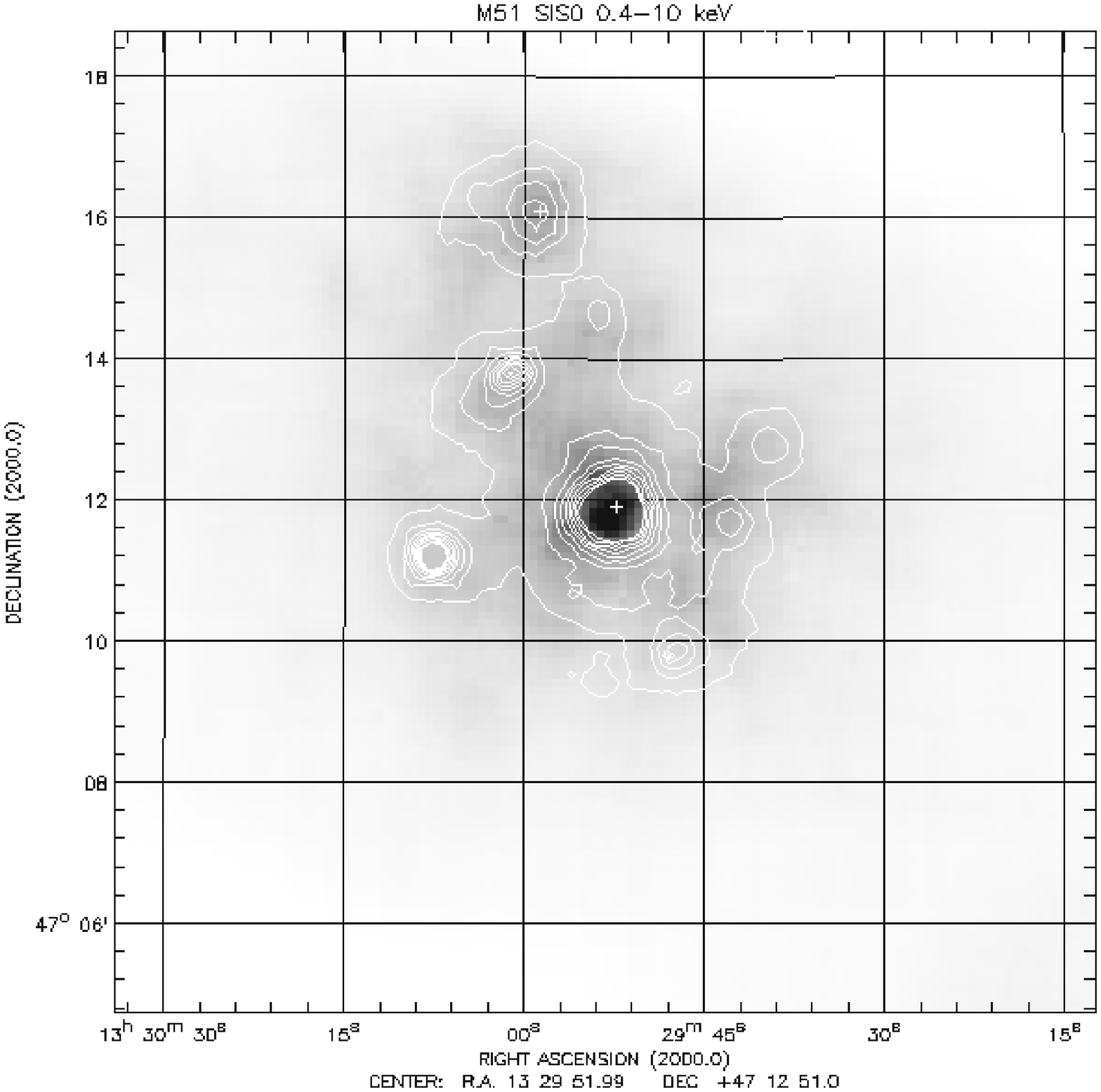}
{3.0in}{0}{52}{35}{-160}{0}
%\plotfiddle{/local/home/neutron/ptak/ngc253/asca/sis0/n253_s0_pspc_cont_15nov95.ps}
\plotfiddle{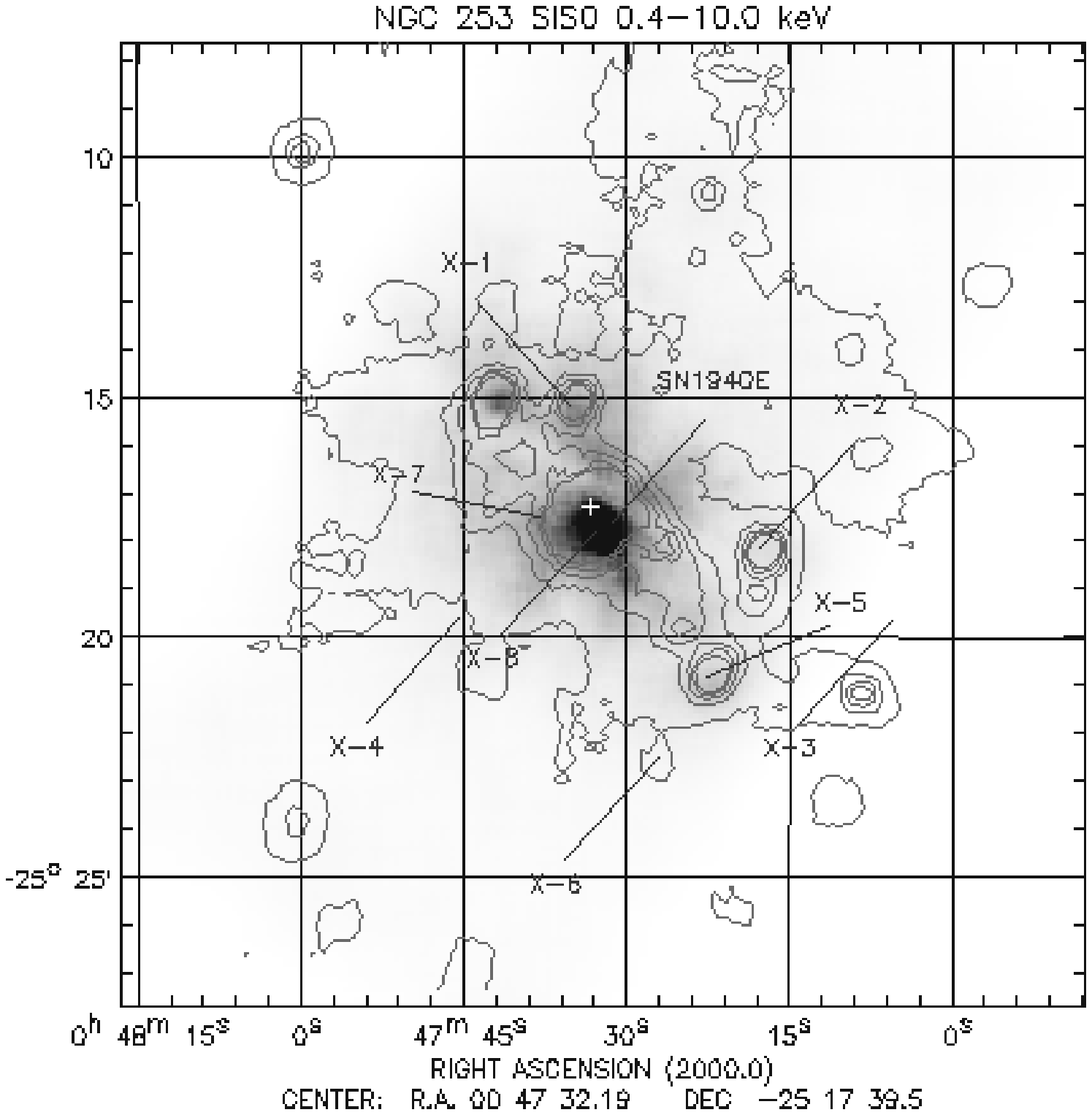}
{2.5in}{0}{71}{48}{-220}{-80}
\caption{\small M51 (top) and NGC 253 (bottom) \asca S0 0.4-10.0 keV images 
with \rosat \pspc 0.1-2.4 keV
contours, adaptively smoothed to a minimum of SNR of $\sim 7$ 
(without background subtraction).  The contours are linearly spaced
from 2 to 20 smoothed counts.  The crosses show the locations of the optical 
nuclei given in NED for NGC 253, M51 (NGC 5194) and its companion (NGC 5195). Although these observations were
performed in 4-CCD mode, only chips 1 and 2 have a significant number of
source counts (note that the dark lanes in the S0 images are due to a gap
between the chips).  The M51 \rosat positions were shifted as described in 
Marston \etal (1995) 
and the \asca image was registered with the \rosat image (using 
only 0.5-2.0 keV photons).  The NGC 253 images have not been shifted.
For distances of 14 Mpc (M51) and 2.5 Mpc (NGC
253), 1' = 4.0 and 0.7 kpc, respectively. \normalsize} 
\end{figure}

\begin{figure}[htbp]
%\plotfiddle{/home/neutron/ptak/m82/asca/sis0/m82_s0_pspc_13nov95.ps}
\plotfiddle{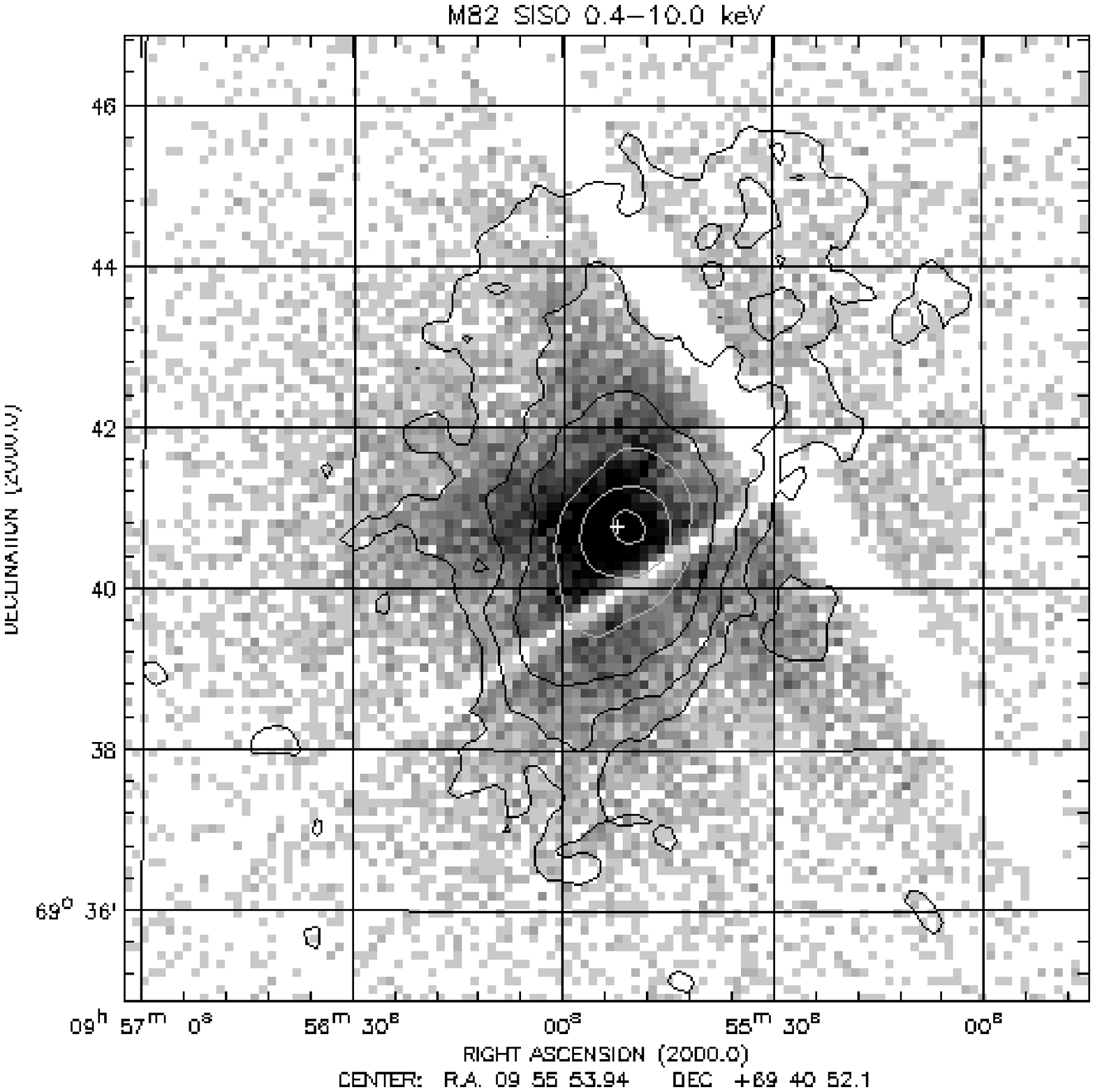}
{2.7in}{0}{58}{40}{-173}{-60}
\caption{\small M82 \asca S0 0.4-10.0 keV image with \rosat \pspc 0.1-2.4 
keV 
contours.  The \rosat image has been smoothed with a 12.5'' Gaussian and the
contours are logarithmically spaced.  The gaps in the S0 image are due to
spacings between the CCD chips.  The cross shows the location of the optical
nucleus.  For a distance of 3.6 Mpc, $1'$ = 1 kpc.
\normalsize}
\end{figure}

For most of the galaxies in our sample, the hard (i.e., E $> 2$ keV) flux
is unresolved while the soft flux is extended over a region greater than
$\sim 1'$ in extent (see, e.g., Makishima \etal 1994; Tsuru \etal 1994; 
Terashima  \etal 1994).  Most of the galaxies have multiple point sources 
resolved by
\rosat while multiple sources are resolved by \asca in only about half of
the galaxies in our sample (see Table 1): NGC 1313 (Petre \etal 1994),
NGC 3628 (Yaqoob \etal 1995a), M33 (Takano \etal 1994), 
NGC 253, M81 (Ishisaki \etal 1996), and possibly
NGC 4258 (Makishima \etal 1994).  Note that the spatial resolution of
the \rosat PSPC is a factor of $\sim 6$ better than the \asca SIS.  
Figures 1 and 2 show the 0.4-10.0 keV \asca S0 images of M51, 
NGC 253, and M82, overlaid with \rosat PSPC (0.1-2.4 keV) 
contours. Multiple sources are clearly visible in both NGC 253 and M51 
(most notably M51's companion), whereas both the \asca and \rosat 
images of
M82 are dominated by a nuclear source with diffuse extended emission (most
visible in a NW plume spanning $\sim 5' = 5$ kpc).
Some of the point sources in NGC 253 are clearly transient.  In all of the
galaxies, including the starbursts, the brightest source is typically the one 
in closest proximity to the optical nucleus. 

\section{Spectral Fitting: A Canonical Model}

\begin{table}[tbp]
\small
\begin{center}
\caption{Absorbed Power-law + Raymond-Smith Fits to \asca Spectra}
\begin{tabular}{lccccccc}
\tableline
Galaxy & Det. & $\rm  N_{H,Gal}^*$ & kT (keV)
& $\rm A/A_{\sun}$ &
$\rm  N_{H,PL}^*$ & $\Gamma$ & $\chi^2$(dof) \\
\tableline
M33$^a$ & S0-S1 & 3.3 & & & & 2.20 & 311(157) \\ \vspace{2pt}
M33$^a$ & G2-G3 & 5.2 & & & & 2.49 & 431(173) \\ \vspace{2pt}
M51 & S0-S1 & $0.16^{\dagger}$ & $0.64^{+0.04}_{-0.05}$ 
& $0.04^{+0.01}_{-0.01}$ &
$25^{+18}_{-20}$ & $1.76^{+1.26}_{-0.98}$ & 194.1(147) \\ \vspace{2pt}
M51 & S0-G3 & $0.16^{\dagger}$ & $0.63^{+0.04}_{-0.04}$ 
& $0.04^{+0.02}_{-0.01}$ 
& $17^{+15}_{-12}$ & $1.52^{+0.61}_{-0.51}$ & 393.8(344) \\ \vspace{2pt}
M81 & See \S 6 \\ \vspace{2pt}
M82 & S0-S1 & 2.2 & 0.78 & 0.06 & 11.3 & 1.65 & 615.4(391) \\ \vspace{2pt}
M82 & S0-G3 & $2.3^{+0.4}_{-0.5}$ & $0.78^{+0.03}_{-0.04}$ & $0.06^{+0.03}_
{-0.02}$ & $11.0^{+4.5}_{-4.3}$ & $1.66^{+0.10}_{-0.09}$ & 1448(1242) \\ \vspace{2pt}
N253 & S0-S1 & $0.13^{\dagger}$ & $0.69^{+0.11}_{-0.07}$ & $1.0^{\dagger}$ 
& $1.1^{+0.4}_{-0.4}$ & $1.82^{+0.10}_{-0.11}$ & 313.4(256) \\ \vspace{2pt}  
N253 & S0-G3 & $0.13^{\dagger}$ & $0.69^{+0.08}_{-0.07}$ & $1.0^{\dagger}$ & 
$1.4^{+0.3}_{-0.3}$ & $1.86^{+0.07}_{-0.06}$ & 834.7 (747) \\ \vspace{2pt}
N1313$^b$ & S0-S1 & $1.2^{+0.4}_{-0.4}$ & & & & $1.8^{+0.1}_{-0.1}$ 
& 93(90) \\ \vspace{2pt} 
N1313$^b$ & G2-G3 & $1.2^{+1.0}_{-0.9}$ & & & & $1.8^{+0.1}_{-0.1}$ & 91(98) \\ \vspace{2pt}
N3079 & S0-S1 & $0.3^{\dagger}$ & $0.77^{+0.09}_{-0.16}$ 
& $0.05^{+0.05}_{-0.03}$ 
& $16^{+24}_{-16}$ & $1.44^{+0.49}_{-1.17}$ & 178.6(184) \\ \vspace{2pt}
N3079 & S0-G3 & $0.3^{\dagger}$ & $0.76^{+0.09}_{-0.16}$ 
& $0.04^{+0.05}_{-0.03}$ & 
$16^{+27}_{-13}$ & $1.76^{+1.02}_{-0.89}$ & 358.0(371) \\ \vspace{2pt}
N3147$^c$ & S0-S1 & $0.34^{+0.42}_{-0.34}$ & & & & 
$1.74^{+0.15}_{-0.13}$ & 194.0(175) \\ \vspace{2pt}
N3147$^c$ & S0-G3 &  $0.35^{+0.35}_{-0.35}$ & $0.65^{\dagger}$ & 
$1.0^{\dagger}$ & & $1.76^{+0.10}_{-0.10}$ & 504.6(514) \\ \vspace{2pt} 
N3310 & S0-S1 & $1.3^{+0.8}_{-0.7}$ & $0.66^{+0.19}_{-0.43}$ 
& $1.0^{\dagger}$ & & $1.70^{+0.34}_{-0.23}$ & 64.4(65) \\ \vspace{2pt}
N3310 & S0-G3 & $1.3^{+0.8}_{-0.6}$ & $0.68^{+0.18}_{-0.39}$ 
& $1.0^{\dagger}$ & & $1.73^{+0.16}_{-0.19}$ & 143.5(141) \\ \vspace{2pt}
N3628$^d$ & S0-G3 & $10^{+3}_{-3}$ & $0.67^{+0.16}_{-0.33}$ 
& $1.0^{\dagger}$ & & $1.7^{\dagger}$ & 55.9(66) \\ \vspace{2pt}
N3998 & S0-S1 & $1.0^{+0.2}_{-0.1}$ & & & & $1.99^{+0.05}_{-0.05}$ 
& 393.5(378)\\ \vspace{2pt}
N3998 & S0-G3 & $0.12^{\dagger}$ & $0.65^{\dagger}$ & 0.0 
& $1.1^{+1.5}_{-0.4}$ & $1.93^{+0.07}_{-0.07}$ & 1046.1(1020)  \\ \vspace{2pt}
N4258$^e$ & G2-G3 & $0.12^{\dagger}$ & $0.5^{+0.2}_{-0.2}$ & $1.0^{\dagger}$ 
& $150^{+20}_{-20}$ & $1.78^{+0.29}_{-0.29}$ & 101(106) \\ \vspace{2pt}
N4579$^f$ & S0-G3 & $0.34^{+0.05}_{-0.04}$ & & & & $1.87^{+0.09}_{-0.09}$ \\ \vspace{2pt}
N4594 & S0-S1 & $0.38^{\dagger}$ & 0.48$^{+0.23}_{-0.14}$ & $>0.04$
& $2.6^{+4.8}_{-1.6}$ & $1.80^{+0.24}_{-0.19}$ & 169.3(183) \\ \vspace{2pt} 
N4594 & S0-G3 & $0.38^{\dagger}$ & $0.64^{+0.20}_{-0.26}$ & $>0.05$
& $5.5^{+3.1}_{-4.0}$ & $1.96^{+0.21}_{-0.20}$ & 332.4(385) \\ \vspace{2pt}
N6946 & S0-G3 & $2.6^{+0.5}_{-0.5}$ & $0.65^{\dagger}$ 
& $0.01^{+0.07}_{-0.01}$ & $34^{+24}_{-20}$ & $2.92^{+0.89}_{-0.74}$ 
& 461.3(452) \\ \vspace{2pt}
N6946 & S0-G3 & $1.6^{+0.7}_{-0.9}$ & $1.00^{+0.12}_{-0.18}$ 
& $0.05^{+0.12}_{-0.04}$ & $26^{+29}_{-26}$ & $2.00^{\dagger}$ & 449.4(452) \\ 
\tableline \tableline
\end{tabular}
\end{center}
Errors not determined for parameters in poorly fitting models ($\chi^2_\nu > 
1.5$) or parameters that do not reduce $\chi^2$ significantly.  Empty
fields indicate parameter was not included in fit.  
Parenthesis indicate 90\% confidence intervals for 2 parameters. \\
$^* \ 10^{21} \rm \ cm^{-2}$. \quad $\rm  N_{H,PL}$ is
absorbtion in addition to $\rm N_{H,Gal}$ applied to power-law. \\
$^{\dagger}$ frozen parameter \\ 
\vspace{-20pt}
\begin{tabbing}
$^a$ Takano \etal (1994) 
\qquad \quad \= $^b$ Source A from Petre \etal (1994) \\
$^c$ Ptak \etal (1996) 
\> $^d$ Yaqoob \etal (1995a) \\
$^e$ Makishima \etal (1994)
\> $^f$ G. Reichert (private communication) \\
\end{tabbing}
\normalsize
\end{table}

The images in Figures 1-2 suggest that the X-ray emission in this galaxy
sample is complex.  Indeed, while several of the galaxies are fit well by
a simple power-law plus absorber model, 
multiple components are typically required.
In general, all the \asca spectra of the
nuclear sources in each galaxy can be modelled adequately with
a power law (with possible absorption in excess of Galactic) plus
an optically thin thermal Raymond-Smith model (with absorption consistent
with the Galactic value). We will refer to this as the canonical 
model for the X-ray spectra of the nuclear 
region of spiral galaxies. The power-law emission and excess absorption
is generally associated with the compact nuclear source wheres the 
soft emission component is extended. Table 2 shows the results of
spectral fitting with this canonical model. In some cases the soft
component dominates the spectrum, in some the power law dominates,
while in others both components contribute significantly to the
overall spectrum.  Figure 3 shows the differing relative
contributions of the hard and soft spectral components in the LINERS
NGC 4594, NGC 3079, and M51, which can be directly compared, in Figure 4, 
to the
X-ray spectra of a typical Seyfert 1 (NGC 3227), a Seyfert 1.5 (NGC 4151)
and a Seyfert 2 (Mkn 3) galaxy.  It is interesting to note that in most
cases the
best-fitting temperature is consistently in the range $\sim 0.6-0.8$ keV
(simulations show that this is not an artifact of the particular \asca 
bandpass).
The physical origin of this result is not yet satisfactorily
understood (see \S 7). 
The elemental abundances are in many cases significantly 
sub-solar and the power-law photon index is typically $\Gamma \sim 1.7-2.0$.
In a few sources the hard power law is absorbed by a column density
greater than $10^{22} \rm \ cm^{-2}$. In the case of NGC 4258, the
column density exceeds $10^{23} \ \rm \ cm^{-2}$ and the X-ray
spectrum is indistinguishable from typical Seyfert 2 galaxies. 

\begin{figure}[tbp]
%\plotfiddle{/home/neutron/ptak/ngc4594/asca/n4594_s0-3_abspl_rs_unf.ps}
\plotfiddle{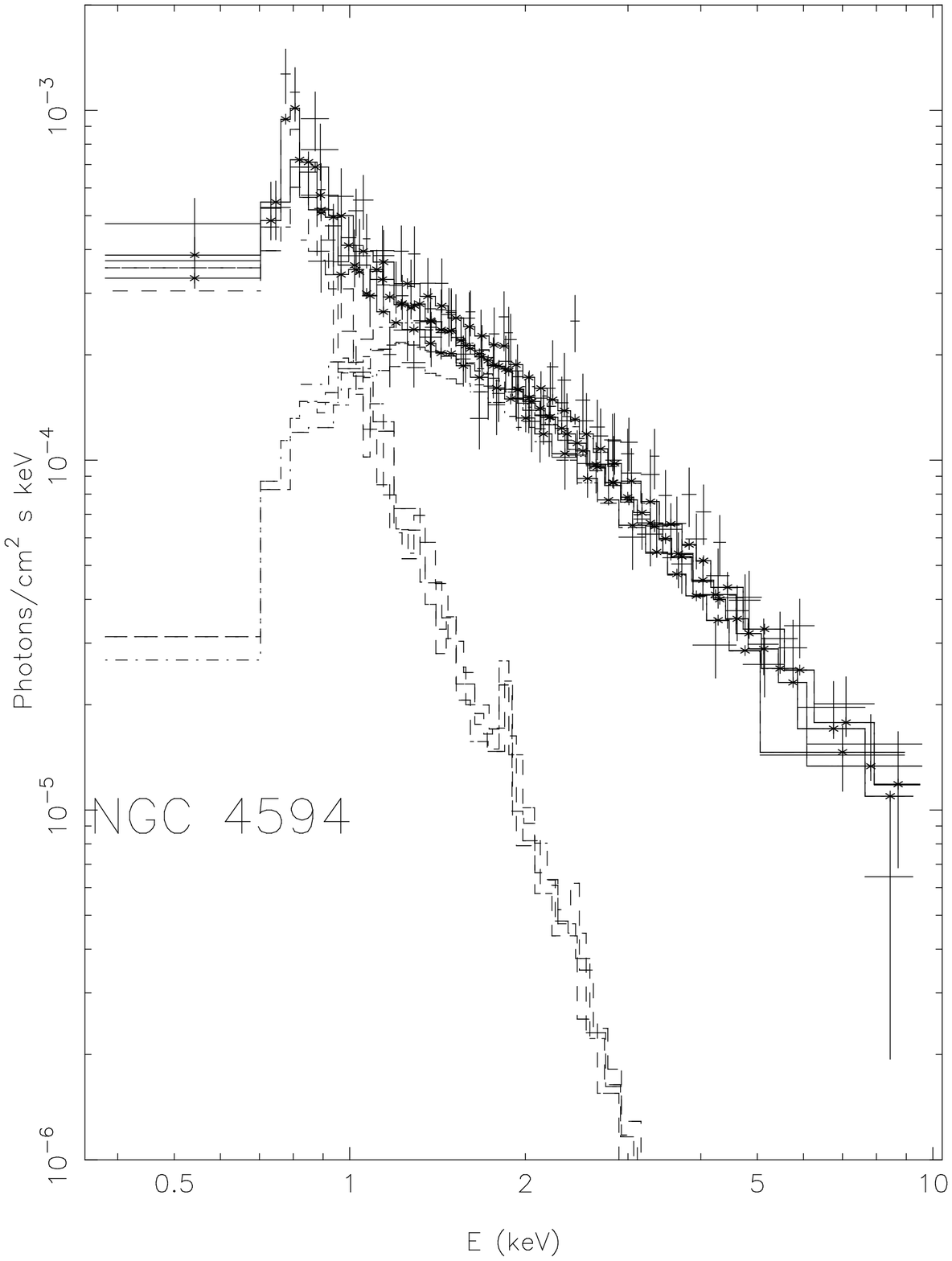}
{2.2in}{0}{65}{25}{-185}{16}
%\plotfiddle{/home/neutron/ptak/ngc3079/asca/n3079_s0-3_abspl_rs_unf.ps}
\plotfiddle{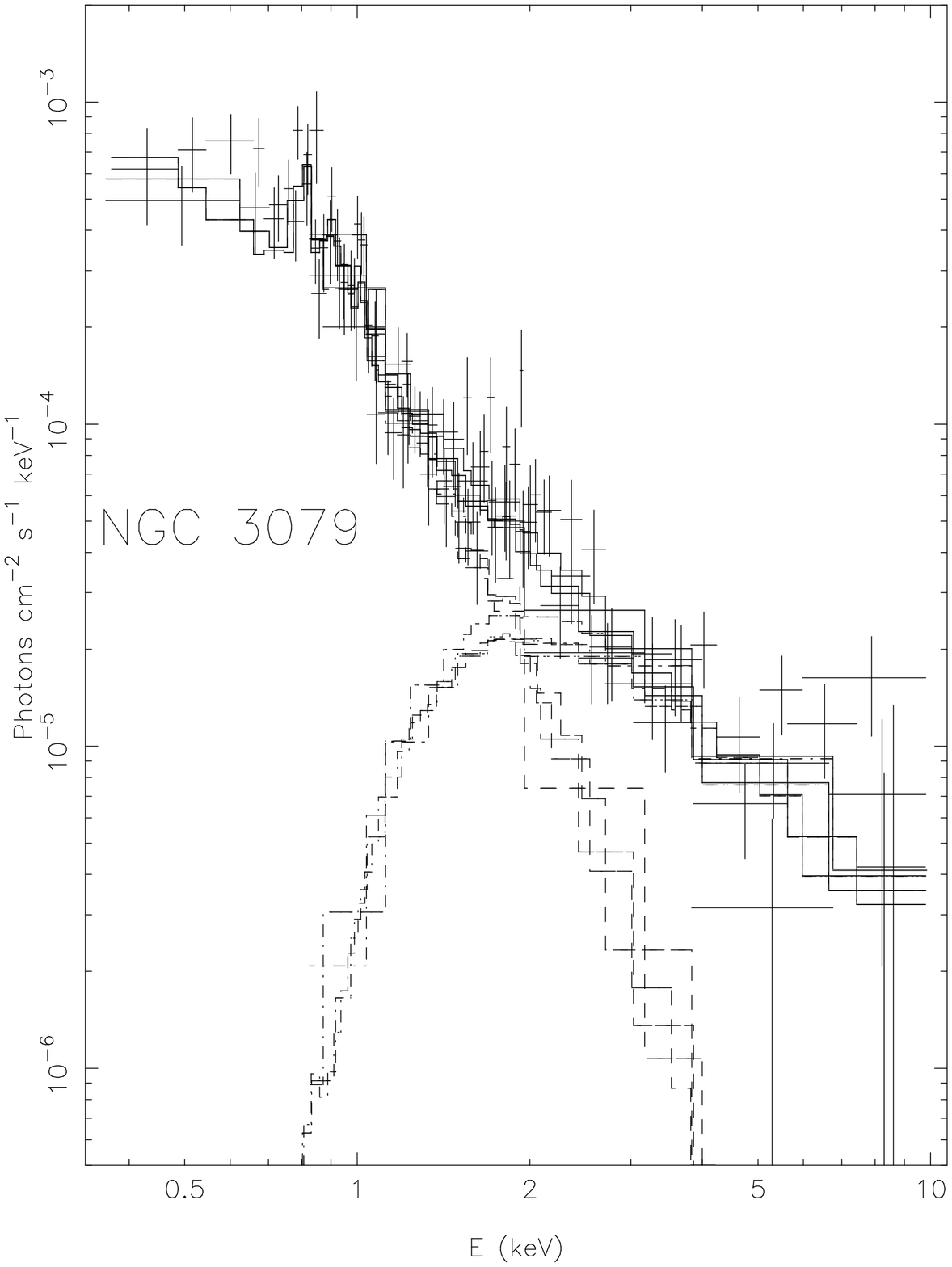}
{2.2in}{0}{65}{25}{-185}{3}
%\plotfiddle{/home/neutron/ptak/m51/asca/m51_s0-3_abspl_rs_unf.ps}
\plotfiddle{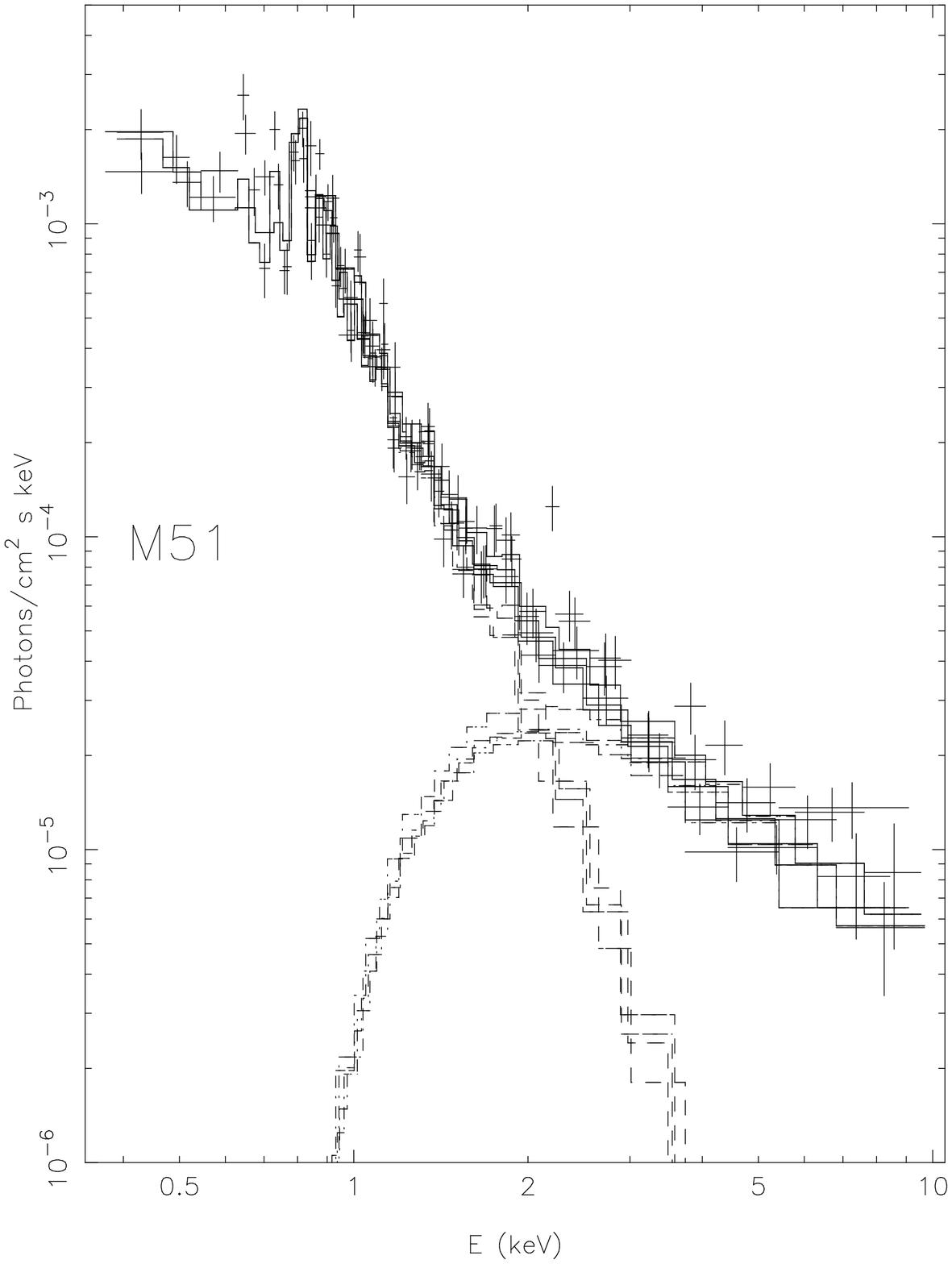}
{2.2in}{0}{65}{25}{-185}{-10}
\caption{\small
The unfolded \asca spectra of M51, NGC 3079 and NGC 4594 (M104) when
fit with the ``canonical'' galaxy model of an absorbed power law + coronal
(Raymond-Smith) plasma.
\normalsize}
\end{figure}

\begin{figure}[tbp]
\plotfiddle{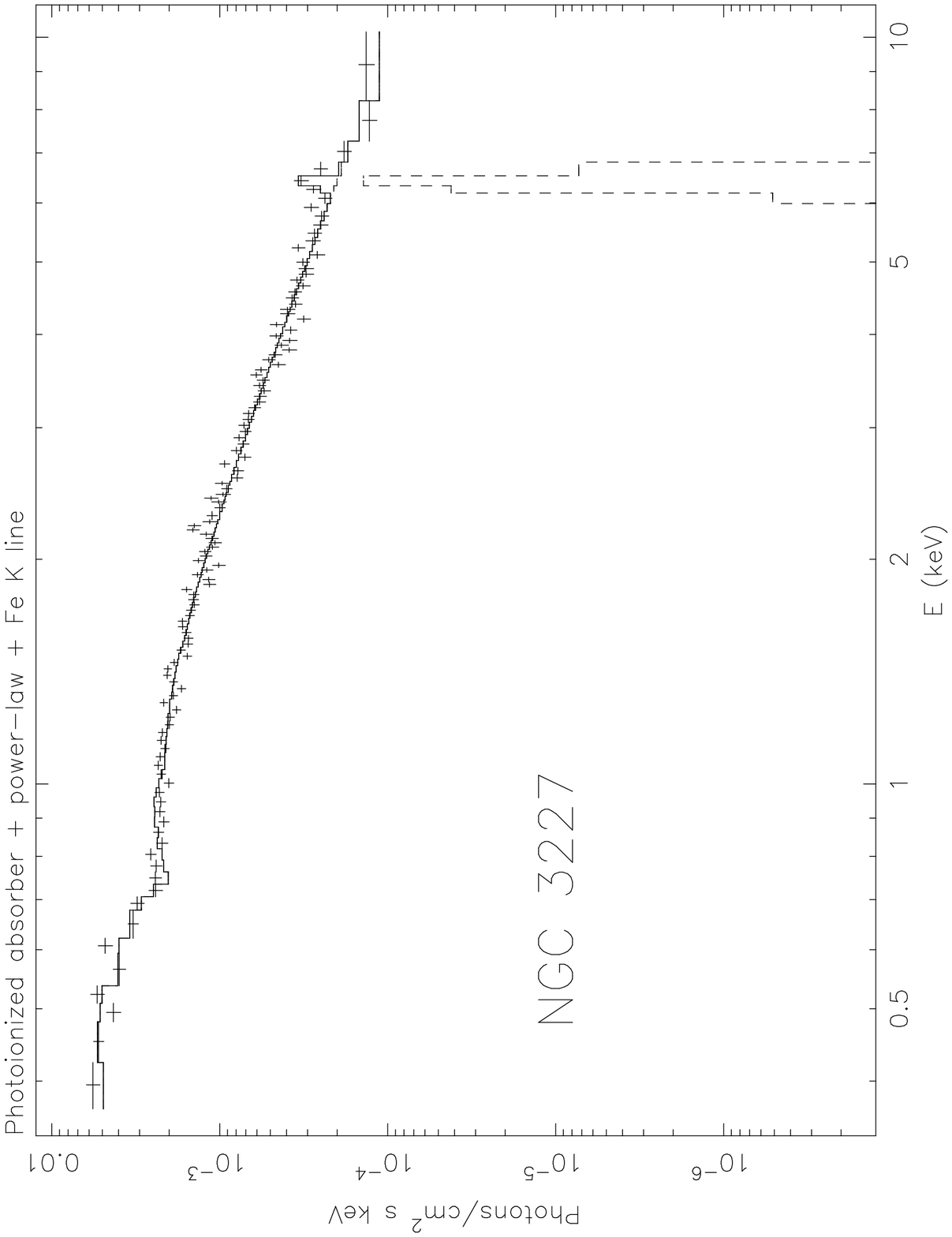}
{2.2in}{270}{48}{30}{-179}{170}
\plotfiddle{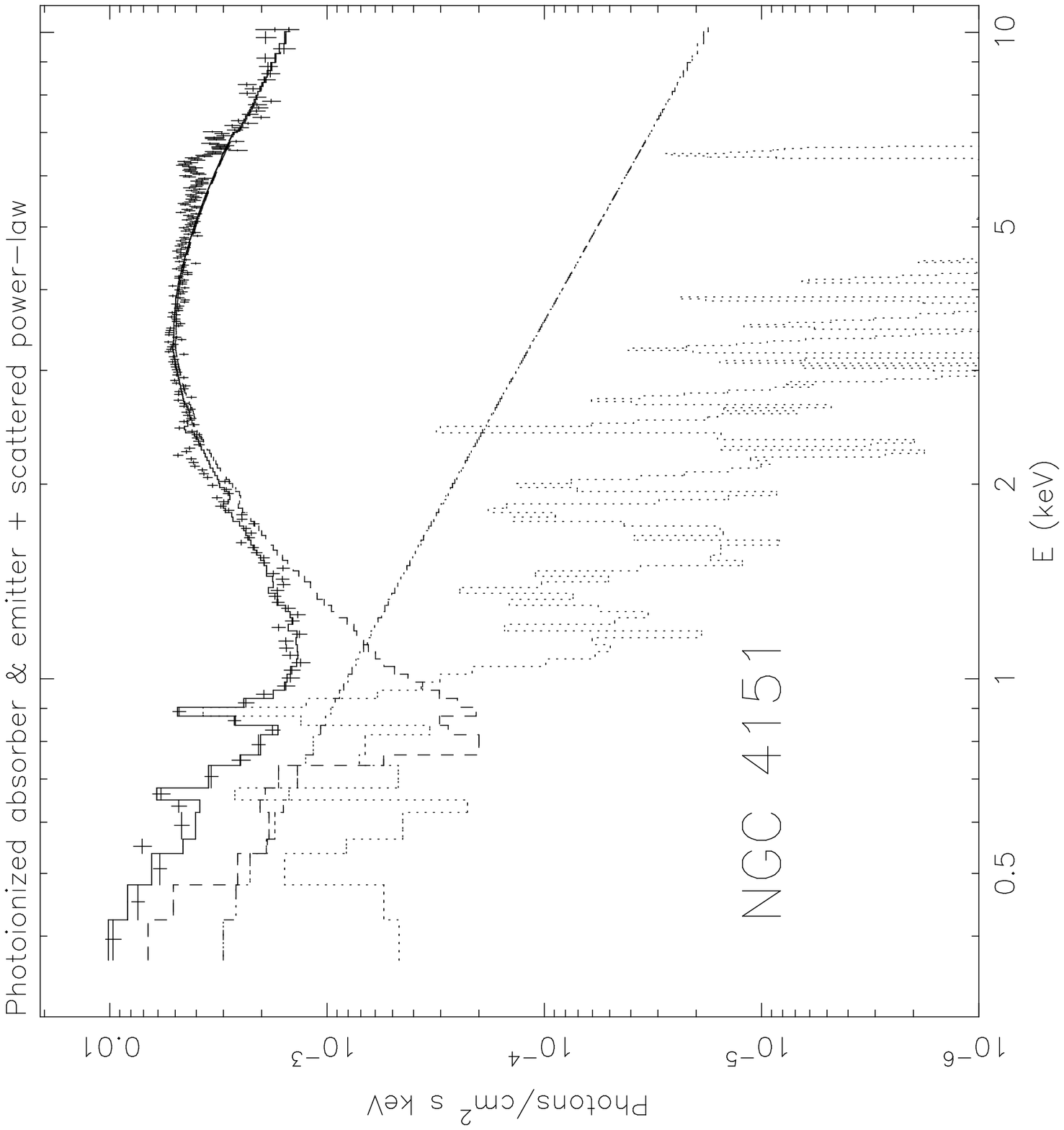}
{2.2in}{270}{60}{30}{-227}{170}
\plotfiddle{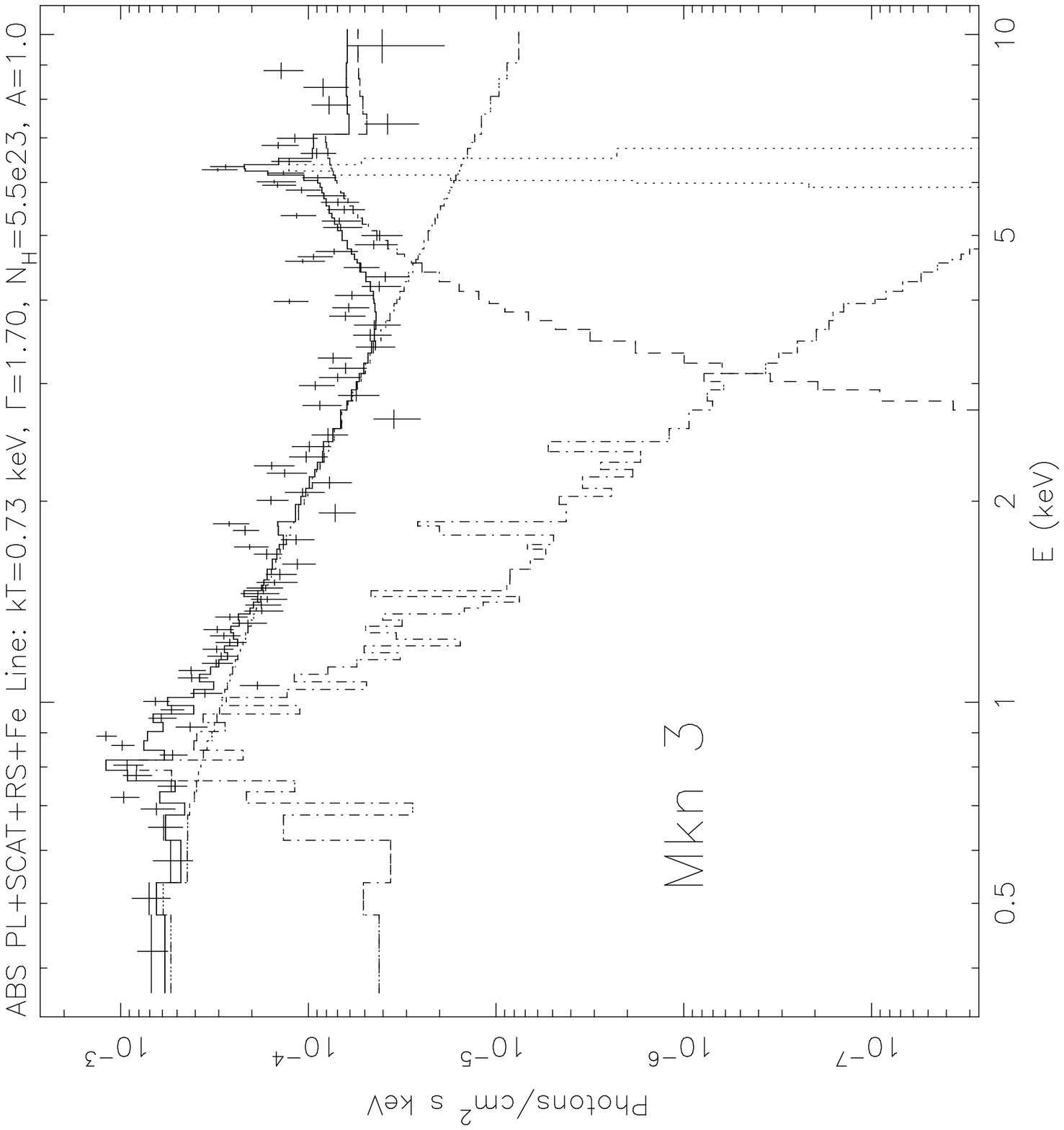}
{2.2in}{270}{60}{30}{-227}{170}
\caption{\small
\asca unfolded S0 spectra of a typical Seyfert 1 (NGC 3227), Seyfert
1.5 (NGC 4151) and a Seyfert 2 galaxy (Mkn 3). For details of the 
best-fitting models see Ptak \etal (1994), Weaver \etal (1994), and
Iwasawa \etal (1994) for NGC 3227, NGC 4151 and Mkn 3 respectively.
\normalsize}
\end{figure}

In some of the galaxies \asca clearly resolves strong
soft X-ray line emission.
Figure 5 illustrates this by 
showing the residuals
obtained from fitting M51, M82, and NGC 253 with an 
absorbed power-law plus a bremsstrahlung model for the soft component. 
The relative line strengths and energies are
potentially powerful diagnostics of the elemental
abundances, temperatures and/or excitation mechanisms of the
emission regions. More detailed analysis will be presented in future
work.

\begin{figure}[tbp]
%\plotfiddle{/home/neutron/ptak/galaxies/m51_s0-1_lines_rat_pretty.ps}
\plotfiddle{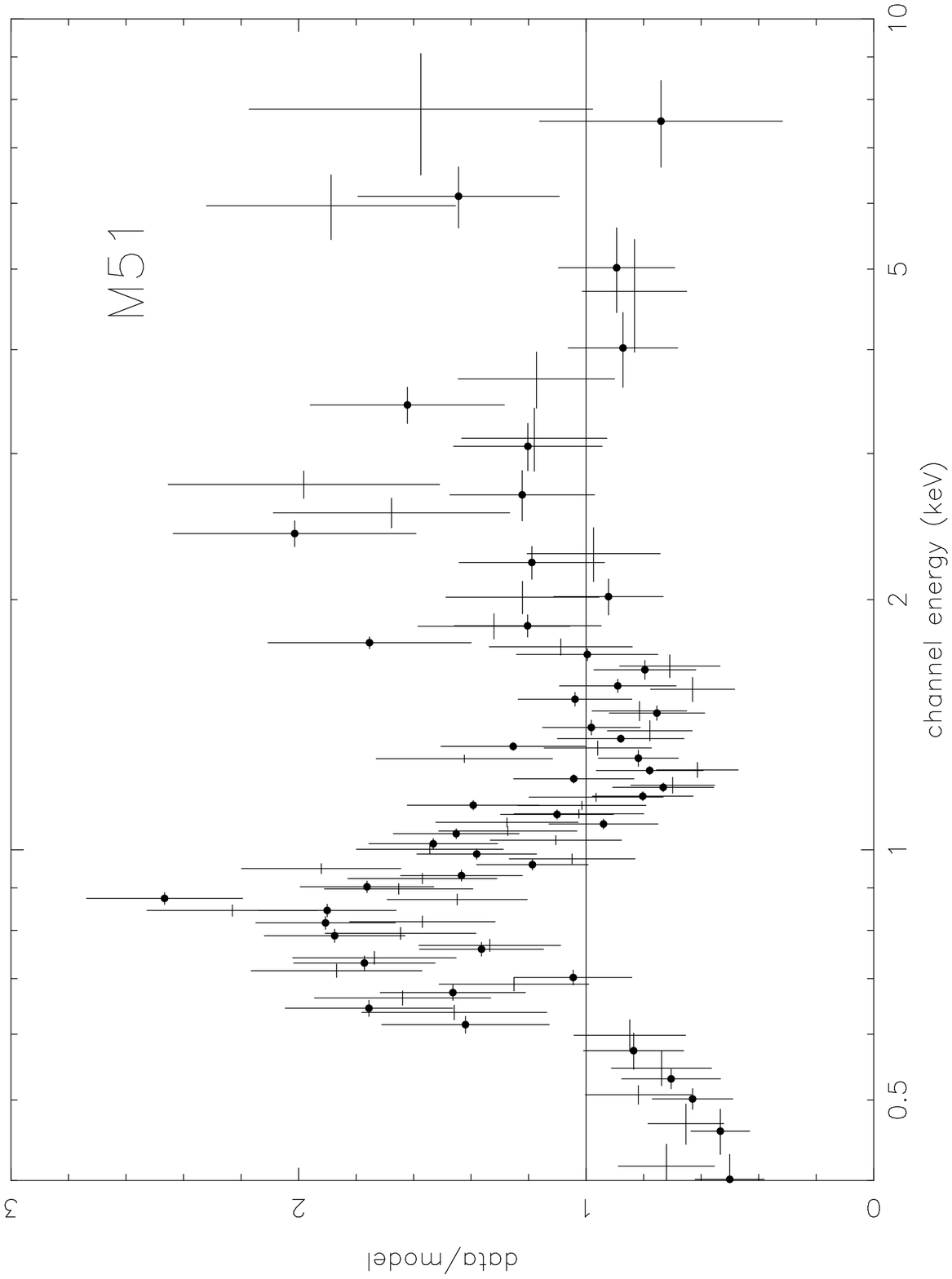}
{2.0in}{270}{50}{30}{-190}{196}
%\plotfiddle{/home/neutron/ptak/galaxies/m82_s0-1_lines_rat_pretty.ps}
\plotfiddle{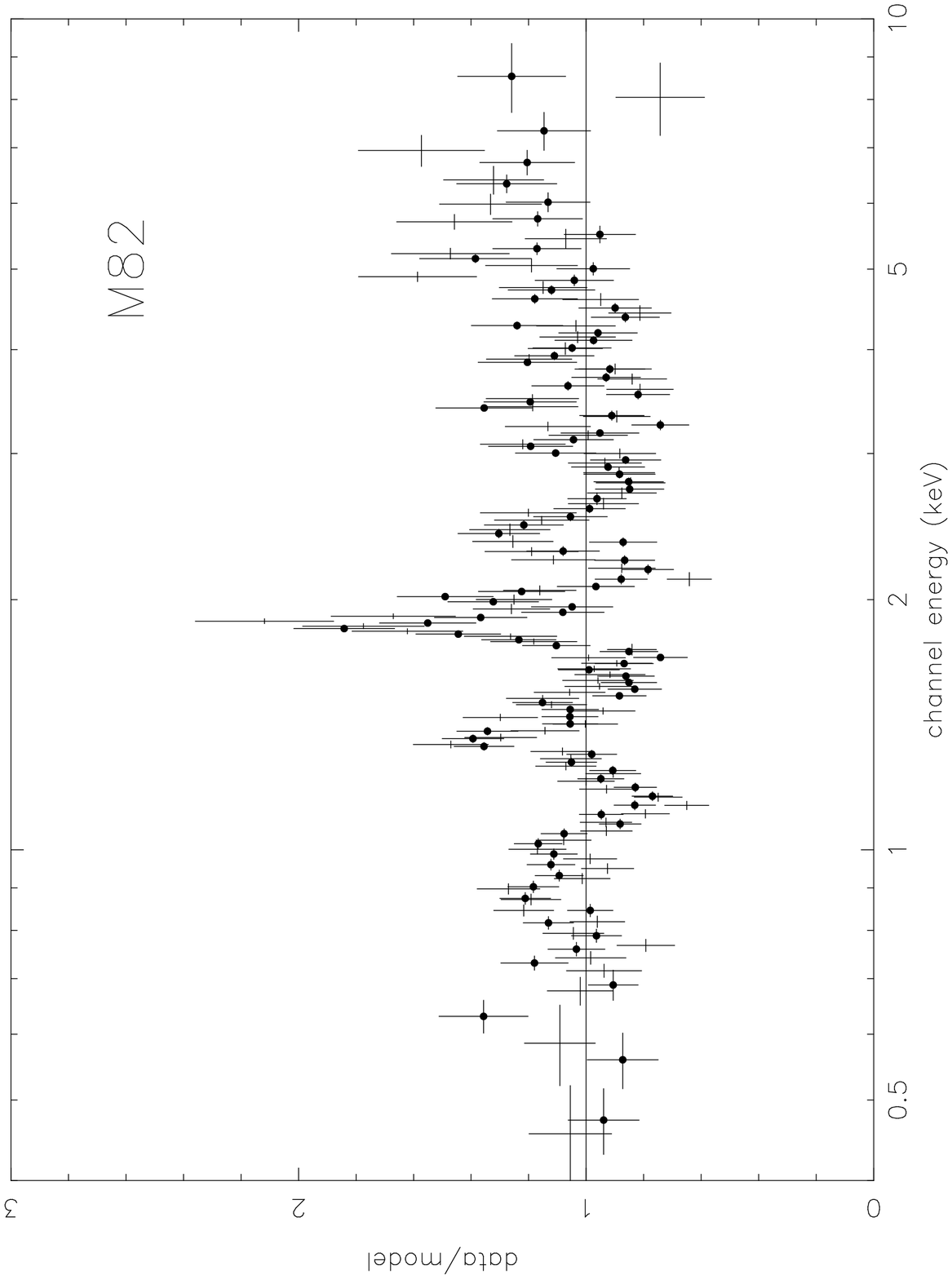}
{2.0in}{270}{50}{30}{-190}{183}
%\plotfiddle{/home/neutron/ptak/galaxies/n253_s0-1_lines_rat_pretty.ps}
\plotfiddle{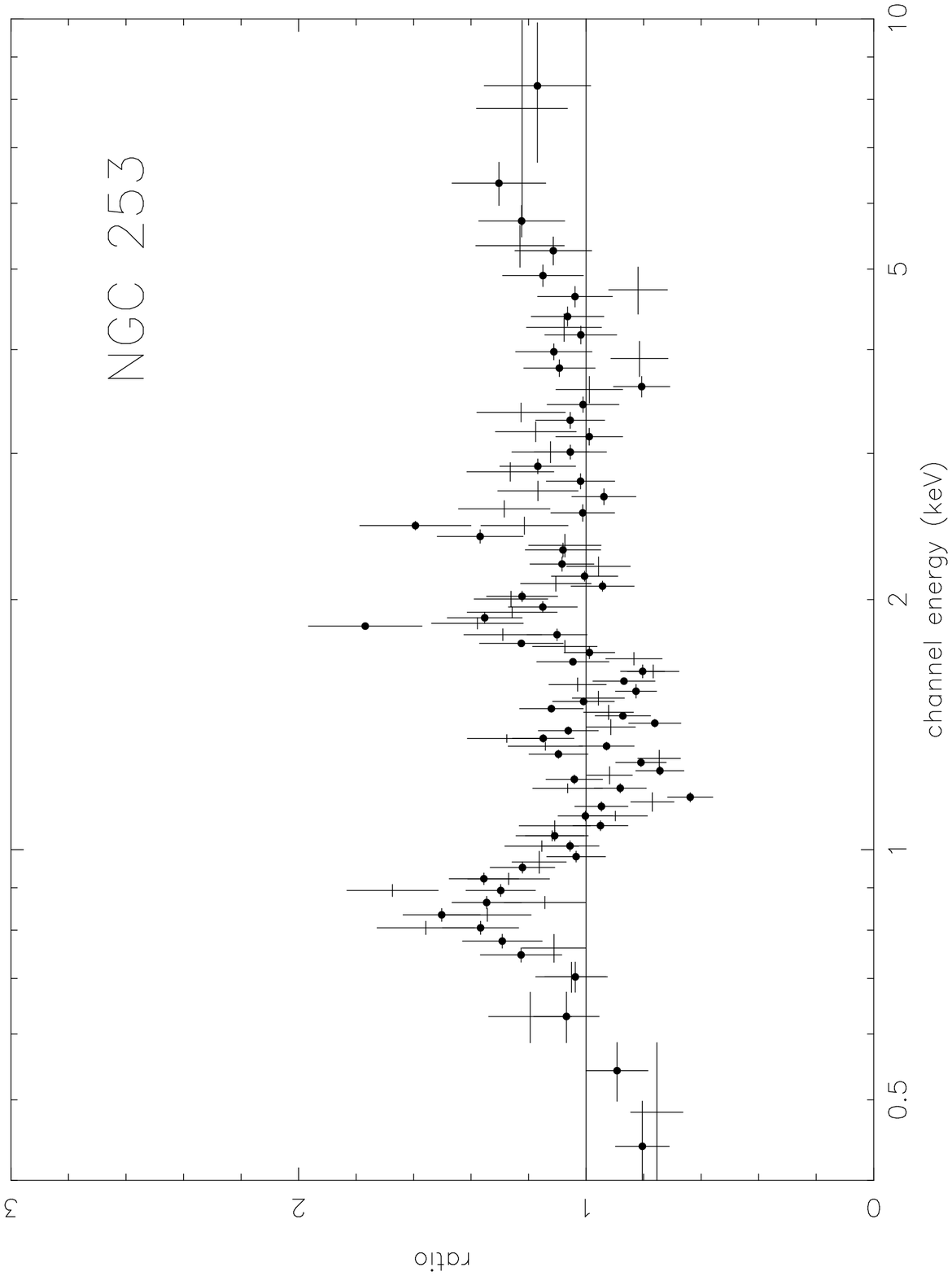}
{2.0in}{270}{50}{30}{-190}{170}
\caption{\small Residuals from fitting with an absorbed power-law + 
bremsstrahlung 
model, with line-like features clearly evident.  Only data from S0 and S1 are
shown for clarity, with S0 points marked with filled circles.  Theses lines 
correspond to 
O VIII (0.65 keV), the Fe-L complex ($\sim 0.8-1.0$ keV), Mg XI-XII 
(1.35, 1.48 keV), Si XIII-XIV (1.87, 2.01 keV), and S XV (2.46 keV).
\normalsize}
\end{figure}
%The differing relative contributions to the of the soft and hard
%spectral components are illustrated for three LINERs, NGC 4594, NGC 3079, 
%and M51 in Figure 4a which shows
%the unfolded spectra. 

Statistically significant Fe K line emission is detected in only
three galaxies: NGC 3147 (see Ptak \etal 1996), NGC 4258
(see Makishima \etal 1994) and M81 (see \S 6), with Fe K emission possibly
detected in M51 (see Terashima \etal 1996). The implications of
these results are discussed in \S 7.

The observed fluxes and luminosities for the sample derived 
from the fit parameters in Table 2 are
shown in Table 3, except for NGC 1313, M33, and NGC 4579 where the 
``canonical'' model 
has not yet been fitted to the data and the fluxes and luminosities were taken
from Petre \etal (1994), Takano \etal (1994), and Reichert (private
communication), respectively. For sources 
in which the power-law component
is absorbed, the intrinsic luminosity can be estimated by multiplying
the observed 2--10 keV (0.5--2 keV) luminosities by the factors
1.01 (1.34),1.10 (5.16) and 1.88 (1168)
for column densities of $10^{21} \ 10^{22}$
and $10^{23} \ \rm cm^{-2}$ respectively (assuming $\Gamma = 2$).
Also shown is the luminosity of the R-S 
component alone in the 0.5-2.0 keV band (excluding NGC 1313, M33, and
NGC 4579).
\small
\begin{table}[htbp]
\begin{center}
\caption{Observed Fluxes and Luminosities} % and Fe-K EW }
\begin{tabular}{lcccccc}
\tableline 
Galaxy & 
$\rm F_{0.5-2.0 \ keV}^{PSPC}$ &
$\rm F_{0.5-2.0 \ keV}^\dagger$ &
$\rm L_{0.5-2.0 \ keV}^\ddagger$ &
$\rm L_{0.5-2.0 \ keV}^{\ddagger,*}$ &
$\rm F_{2-10 \ keV}^\dagger$ &
$\rm L_{2-10 \ keV}^\ddagger$ \\
%& Fe-K E (keV) & Fe-K EW (eV) \\
%& $(\rm 10^{-12} \ \rm ergs \ s^{-1} \ cm^{-2})$ & 
%$(\rm 10^{40} \ \rm ergs \ s^{-1})$ & 
%$(\rm 10^{-12} \ \rm ergs \ s^{-1} \ cm^{-2})$ & 
%$(\rm 10^{40} \rm \ ergs \ s^{-1})$ & 
%(keV) & (eV) \\
\tableline
N3998 & 5.0 & 4.0 & 28. & 3.3 & 6.5 & 46. \\
N4579 & 5.4 & 1.8 & 28 & & 2.9 & 45. \\
N3147 & 1.0 & 0.63 & 24. & 1.5 & 1.2 & 45. \\
N4594 & 1.5 & 0.87 & 5.3 & 2.2 & 1.9 & 12. \\
N3310 & 0.67 & 0.72 & 3.2 & 0.76  & 1.3 & 5.9 \\
N3079 & 0.48 & 0.44 & 4.4 & 0.4 & 0.53 & 5.3 \\
N4258$^a$ & 2.0 & 1.2 & 1.5 & 1.5 & 2.9 & 3.6 \\
M82 & 11. & 7.7 & 1.2 & 0.94 & 18. & 2.8 \\
N3628 & 0.61 & 0.17 & 0.45 & 0.16 & 0.85 & 2.3 \\
M81 & 5.8 & 7.7 & 1.2 & 0.28 & 13.6 & 2.1 \\
M51 & 1.6 & 1.1 & 2.5 & 2.4 & 0.75 & 1.8 \\
N6946 & 1.4 & 1.0 & 0.82 & 0.80 & 0.82 & 0.68 \\
N253 & 3.0 & 2.3 & 0.17 & 0.03 & 4.0 & 0.30 \\
N1313$^b$ & 1.1 & 0.54 & 0.13 & & 1.2 & 0.29 \\
M33$^c$ & 6.3 & 5.7 & 0.10 & & 9.5 & 0.16 \\ 
\tableline
\tableline
\end{tabular}
\end{center}
\footnotesize
$^\dagger$  $(\rm 10^{-12} \ \rm ergs \ s^{-1} \ cm^{-2})$ \\
$^\ddagger$ $(\rm 10^{40} \ \rm ergs \ s^{-1})$ \\
$^*$ Soft component only \\
$^a$ Makishima \etal (1994) \qquad 
$^b$ Source A in Petre \etal (1994) \\
$^c$ Takano \etal (1994) \\
\end{table}
\normalsize

Given the strong evidence for long-term variability (see below) in some of
these galaxies, simultaneously fitting \rosat and \asca spectra that are not
simultaneous may be misleading, particularly if the observed variability is
primarily due to only one component in a multiple-component model.  
Accordingly, the \rosat spectra were fit independently (to be discussed in 
future work), and Table 3 shows the 0.5-2.0 keV flux derived from the best
fitting model. 

\section{Variability}
No short-term (less than 1 day) variability has been observed in our galaxy
sample (see, however, the discussion of M81 below), with upper limits
of typically $\Delta I/I \sim 20\%$ over $\sim 50$ ks in the 0.8-10.0 keV
bandpass (only the GIS data were used to 
search for short-term variability as they 
are less sensitive to pointing stability than the SIS).  However, short-term
variability in the \rosat bandpass has been observed in a point source in M82  
(Collura \etal 1994).  Long-term variability, on the other hand, has been
observed in many of these galaxies (several LINERs in Reichert
\etal 1994; source B in NGC 1313 in Petre \etal 1994; 
NGC 3628 in Yaqoob \etal 1995a and Dahlem \etal 1995).  
Variability is also clearly evident among
the \rosat and \asca 0.5-2.0 keV fluxes for some of the galaxies
listed in Table 3.

\section{The Nuclear Source in M81: A Case Study}

The nearby spiral galaxy M81 is the brightest of the known LINERs and
has been the subject of many detailed investigations.  
Extensive studies in the X-ray band have
been carried out with {\it Einstein} 
(Elvis \& van Speybroeck 1982; Fabbiano 1988),
\ginga (Ohashi \& Tsuru 1992), \rosat (Boller \etal 1992)
and \bbxrt (Petre \etal 1993). Although at least
9 X-ray sources have been resolved using {\it Einstein}
data, the emission is dominated by the nuclear source
(also known as X-5).
One of the motivating forces for studying LINERs and M81 in
particular is that it has often been suggested that LINERs represent
the missing link between nearby `normal' galaxies and AGN.
Studying the X-ray spectra of LINERs and comparison with
the large body of knowledge of the X-ray properties of AGN can
provide powerful diagnoses. The nuclear source in M81 has
been recognized as an active nucleus for some time (Peimbert
\& Torres-Peimbert 1981). Prior to \asca it was known that the
X-ray continuum above $\sim 1$ keV is likely to be due mainly
to a nonthermal component (evidenced by the lack of strong soft  X-ray
line emission), with a photon power-law index of $\Gamma \sim 2.0-2.2$.
Note however that {\it Einstein} measured a much steeper spectrum
(the reason is still unknown). Long-term flux variability has been observed
and even short-term variability (factor of 2 in $\sim 600$s) has been
reported (Mushotzky, private communication).

\asca observed M81 on ten occasions between April 1993 and April 1995,
providing the first high statistical quality, high sensitivity, moderate
energy-resolution, spatially resolved 
hard X-ray spectra of the nuclear source. 
These observations also provide the first opportunity to study hard
X-ray variability free from the ambiguities of non-imaging
data. Here we present preliminary
results using archival data 
combined from three of the observations which were
made with the \asca instruments in similar modes of operation,
between April 16 and May 1 in 1993. 
Ishisaki \etal (1996) have presented results from eight of the ten
observations, three of which are not yet public as of November 1995.
Ishisaki \etal (1996) report the discovery of Fe K line emission
which may have a broad and/or complex structure. However the line
parameters obtained from the SIS and GIS (and indeed the continuum
parameters) are discrepant and one of the goals of our independent
analysis is to investigate this discrepancy. 
Whether there is a single complex line or multiple iron line
components and
whether the line(s) originate in hot or cold matter has
important consequences for models of the structure of the active nucleus.
A more detailed study
resulting from this preliminary work will be published elsewhere.

\subsection{The ASCA Data}

Details of the ten \asca observations of M81 can be found in
Table 1 of Ishisaki \etal (1996). The observations were made
with the SIS in different clocking modes and split thresholds.
Moreover, in all four instruments, X-5 was placed on different
physical positions on the detectors in different observations.
However in observations 3 to 5 (as listed in Ishisaki \etal 1996)
the SIS operation modes were the same, the source was placed
in the same position on each detector and
was at similar intensity levels. Thus, in order to minimize
the introduction of systematic errors we use the combined
data from only these three observations in order to investigate
the spectral shape and the iron line emission. We combined SIS data
taken in FAINT and BRIGHT modes and did {\it not} correct for the
so-called Dark Frame Error (DFE). This may broaden the energy
resolution of the SIS somewhat and may induce energy shifts of
the order $\sim 10-30$ eV. We obtained four spectra with exposure
times in the range 99--124 Ks (after cleaning).

Note that X-5 was {\it not}
positioned on one of the nominal SIS chips 
in these three observations (i.e. chip 1 for S0 and
chip 3 for S1). 
The SIS calibration is less certain for
the non-standard chips since almost all of the point-source \asca
observations are made on the nominal chips. 
Individual spectra were extracted from a circle
of radius 2' for the SIS and 3' for the GIS. As estimated by
Ishisaki \etal (1996), even for a 3' circle the contamination
from the nearby supernova, SN1993J, and the
source X-6, should be less than 10\%.
Background spectra were obtained from source-free parts of the
detectors (the integration times being large enough to compensate
for the small areas of the background regions).

%\begin{table}[htbp]

%\begin{tabular}{lccccccc}

%Obs & Date & Exp. $^{a}$ & $\Gamma$ & $N_{H}$ & 
%$F_{0.5-2.0 \rm keV}^{b}$ & $F_{2-10 \rm keV}^{b}$ & $\chi^{2}$ (dof) \\
%& & & & & & & \\
%& (1993) & (Ks) & & ($10^{20} \rm \ cm^{-2}$ & & & \\
%& & & & & & & \\
%1 & 5 Apr & & & & & & \\
%& & & & & & & \\
%2 & 7 Apr & & & & & & \\
%& & & & & & & \\
%3 & 16 Apr & & & & & & \\
%& & & & & & & \\
%4 & 25 Apr & & & & & & \\
%& & & & & & & \\
%5 & 1 May & & & & & & \\
%& & & & & & & \\
%6 & 19 May & & & & & & \\
%& & & & & & & \\
%7 & 24 Oct & & & & & & \\
%& & & & & & & \\
%\end{tabular}

%$^{a}$ Mean exposure time for four instruments. \\
%$^{b}$ Mean fluxes in units of $10^{-11} \rm \ erg \ cm^{-2} \ s^{-1}$,
%averaged over four instruments. \\

%\end{table}

\subsection{X-ray Variability}
\begin{figure}[htbp]
\plotfiddle{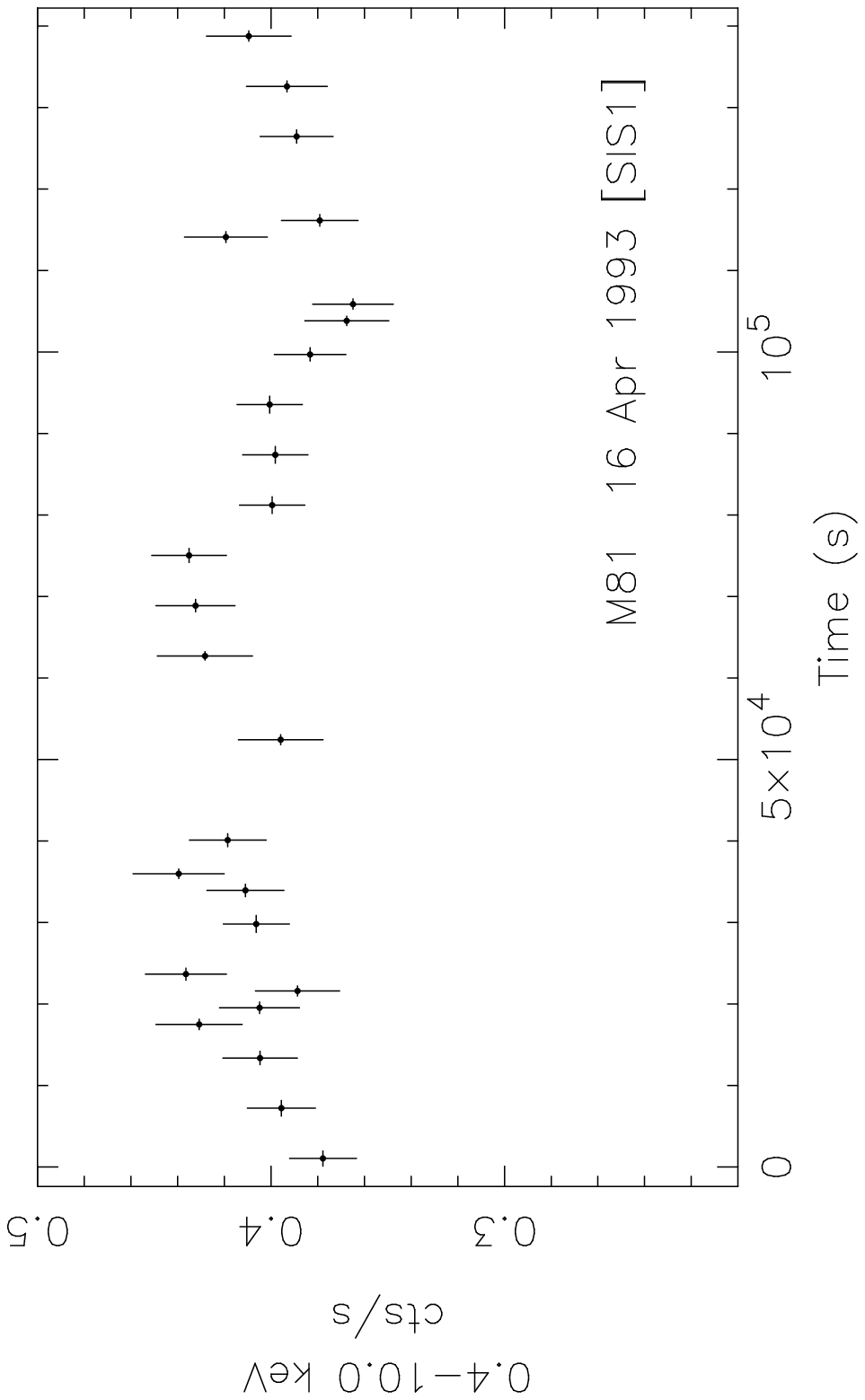}{2.0in}{270}{75}{50}{-280}{200}
\caption{\small The S1 light curve of the extended observation \normalsize}
of M81 during 1993, April 16--18. The data are binned at 2056 s.
Note that the source is not close to the edges of the chip reducing the
effects of positional instability on the light curve.
\end{figure}
Figure 6 shows the 0.4--10 keV SIS1 light curve from the
longest of the ten ASCA observations which spanned almost 2 days
between April 16-18 1993. SIS1 yields the highest count-rate among
the four instruments in this case. Variability of the
order of $\sim 20\%$ on a timescale of hours is evident and this
is typical of this source (see also Ishisaki \etal 1996).
Among the eight observations analysed by Ishisaki \etal (1996),
long term variability by a factor of up to $\sim 1.7$ was found
corresponding to a 2--10 keV flux in the range
1.4--2.4 $\times 10^{-11} \rm \ erg \ cm^{-2} \ s^{-1}$
and 2--10 keV luminosity in the range $4.2-7.4 \times 10^{40}
\rm \ erg s^{-1}$ ($H_{0} = 50 \rm \ km \ s^{-1} \ Mpc^{-1}$ is
assumed throughout).

\subsection{Spectral Fitting}
\begin{figure}[htbp]
\plotfiddle{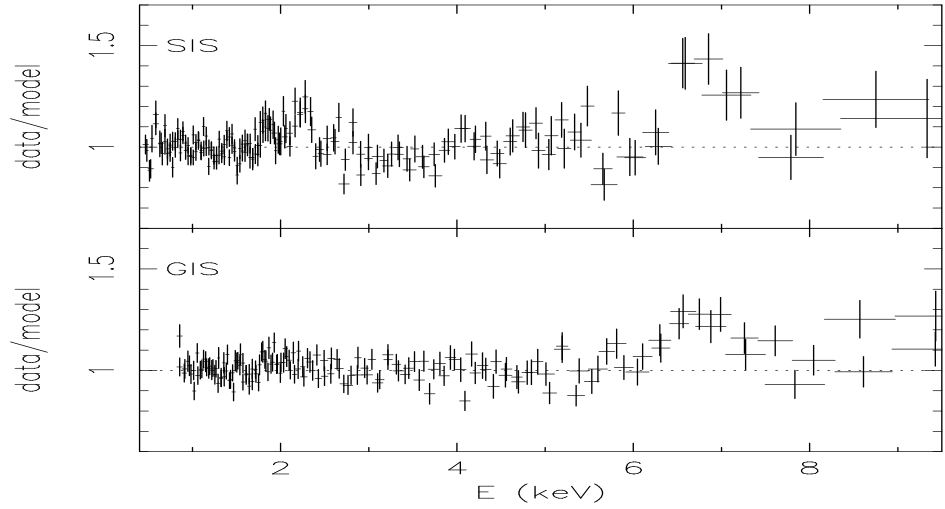}{2.5in}{270}{100}{50}{-380}{250}
\caption{\small Ratio of M81 \asca data to best-fitting power-law plus
absorber model, showing the iron K line residuals. \normalsize}
\end{figure}
Figure 7 shows the result of fitting the combined data from
observations 3--5 with a simple power law plus absorber model.
Due to the different properties of the SIS and GIS instruments,
simultaneous fits are performed with S0 and S1 and then repeated
with G2 and G3 together.
A highly significant iron emission line feature is evident
in both SIS and GIS data. In the SIS two line-like features
are also seen at soft X-ray energies but these may be, at least
in part, instrumental features. They will subsequently be modelled
with simple Gaussians. The iron K feature is also first modelled with
a single Gaussian. The results are shown in Table 4 and
\begin{figure}[htbp]
\plotfiddle{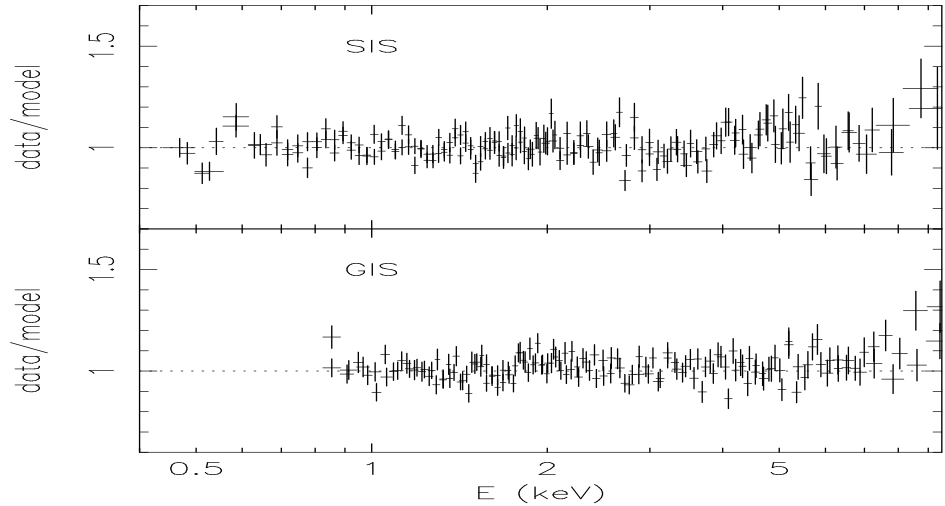}{2.5in}{270}{100}{50}{-380}{250}
\caption{\small Ratio of M81 \asca data to best-fitting power-law
plus absorber model, including a single Gaussian model for the iron
K line. \normalsize}
\end{figure}
Figure 8 shows the ratio of data to best-fitting model.
The photon index, $\Gamma$, is different at the 90\% confidence 
level for the SIS and GIS. This may be due in part to the larger integration
region and spatial resolution of the GIS, leading to greater contamination.
Also, the GIS is not sensitive to such small column densities 
($\Gamma$ and $N_{H}$ are correlated parameters). The SIS column
is consistent with the Galactic value of $4.3 \times 10^{20} \rm
\ cm^{-2}$.

\begin{table}[htbp]
\begin{center}
\caption{Power Law Plus Absorber and Single Gaussian Model}
\begin{tabular}{lccccccc}
\tableline 
DET  & $\Gamma$ & $N_{H}$ & $E_{\rm Fe}$ & $\sigma_{\rm Fe}$ &
$I_{\rm Fe} ^{a}$ & EW & $\chi^{2}$  (dof) \\

& & ($10^{20} \rm \ cm^{-2}$) & (keV) & (keV) & & (eV) & \\

& & & & & & & \\

S0,S1 & $1.91^{+0.04}_{-0.05}$ & $6.5^{+0.13}_{-0.13}$ &
$6.78^{+0.24}_{-0.22}$ & $0.30^{+0.26}_{-0.17}$ & $4.2^{+2.5}_{-2.3}$
& $385^{+229}_{-211}$ & 476.5 (412) \\
& & & & & & & \\
S2,S3 & $1.72^{+0.05}_{-0.03}$ & $0.0^{+4.4}_{-0.0}$ & 
$6.68^{+0.19}_{-0.18}$ & $0.21^{+0.34}_{-0.21}$ & $3.0^{+2.1}_{-1.6}$
& $228^{+160}_{-122}$ & 1023 (970) \\
& & & & & & & \\
\tableline
\tableline
\end{tabular}
\end{center}
\footnotesize
Errors are 90\% confidence for 5 interesting parameters. \\
$^{a}$ Iron line intensity in units of $10^{-5} \rm \ photons \ cm^{-2} \
s^{-1}$. \\
\end{table}
\normalsize
The SIS and GIS iron line parameters {\it are} consistent with each
other. See Table 4 and
\begin{figure}
\plotfiddle{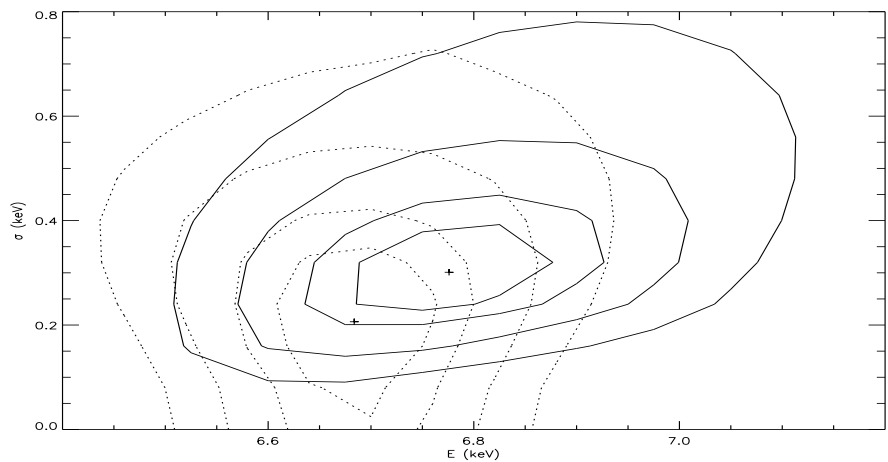}{2.5in}{0}{70}{50}{-215}{-190}
\caption{\small Confidence contours of the center energy and intrinsic
width of the iron K line in M81 when it is modelled with a single
Gaussian.   The contour levels for the SIS data (solid lines) and the GIS 
data (dotted lines) correspond to 68\%, 90\% and 99\% for 2
interesting parameters and the final contour to 99\% for 5 interesting
parameters ($\Delta \chi^{2} = 2.279$, 4.605, 9.21
and 15.086 respectively). Note that the third contour is approximately equal to
90\% confidence for 5 interesting parameters
($\Delta \chi^{2} = 9.24$). \normalsize}
\end{figure}
Figure 9 which shows confidence contours of the line energy 
($E_{\rm Fe}$) and intrinsic Gaussian width ($\sigma_{\rm Fe}$).
The line is required to be broader than the SIS resolution at a
high level of confidence. The best-fitting SIS width corresponds
to $\sim 0.71$ keV FWHM or $\sim 30,000 \rm \ km \ s^{-1}$ FWHM.
The line energy is consistent with an origin in He-like iron.
The equivalent width (EW) of the line is large, $\sim 200-400$ eV
and may be problematic for models of the line emission (see \S 6.4).

The 0.5--2 keV and 2--10 keV fluxes averaged over four instruments
are 7.7 and $13.6 \times 10^{-12} \rm \ erg \ cm^{-2} \ s^{-1}$ 
respectively, corresponding to luminosities of 2.4 and 
$4.2 \times 10^{40} \rm \ erg \ s^{-1}$ respectively. In order
to test for the presence of an additional soft component we
added a Raymond-Smith component and obtained a reduction in
$\chi^{2}$ of 14 (SIS) and 3 (GIS) for four additional parameters
(two normalizations, temperature and metal abundance). For the
SIS we obtained $KT \sim 1.25$ keV and negligible metal abundance
($<0.01$ relative to cosmic). Thus, the case for an additional
soft continuum component is not strong. Note that thermal models
of the {\it hard} continuum are very strongly ruled out.

\subsection{Origin of the Iron K Emission}

In order to investigate the origin of the iron K line we
performed additional spectral fits using the SIS 
data as follows. \\
{\bf 1. Multiple Line Complex} To test the possibility that 
a multiple set of narrow lines may be mimicking a
a broad line, we replaced the single broad Gaussian
in \S 6.3 with three narrow Gaussians (intrinsic width of each
fixed at 0.1 keV). The best-fitting energies are 6.52, 6.76, and
7.15 keV with corresponding equivalent widths of 95, 65 and 178 eV 
respectively. We get $\chi^{2} = 499.8$ (409 d.o.f.) which,
even with three more free parameters, is higher
by 23.3  than the $\chi^{2}$ for 
the single broad Gaussian model. If the intrinsic widths of 
the lines are allowed to be free, two of the lines are forced
to zero intensity and the single broad Gaussian solution is once
again recovered. \\
{\bf 2. Thermal Comptonization} We replaced the single broad
Gaussian in \S 6.3 with a model of a monochromatic line
broadened by Compton scattering in plasma with temperature 
$kT$ and optical depth $\tau$ (both free parameters).
The best-fitting values are $kT =13.1$ keV and $\tau = 0.15$
but $\chi^{2} = 508$ (412 d.o.f.) which is 31.5 higher than
the single broad Gaussian model, yet the number of free parameters
is the same in each case. \\
{\bf 3. Relativistic Disk Model} We also tried a relativistic disk
model for the iron line profile. See, for example Tanaka \etal (1995)
and Yaqoob \etal (1995b) where the model has been applied to the
iron line profiles in the Seyfert galaxies MCG $-$6$-$30$-$15 and
NGC 4151 respectively. The best-fitting rest-frame line energy
is 6.73 keV but the 90\% confidence errors
allow any value in the range 6.4--6.9 keV so
it was fixed at its best-fitting value.
We obtained $\chi^{2} = 480$ (412 d.o.f.)
and a disk inclination angle of $40^{+10}_{-12}$ degrees and
an EW of $523 \pm 245$ eV. It is difficult to produce such a 
large equivalent width without increasing the iron abundance 
significantly above solar. The same problem has been encountered
with the broad lines detected in some classical Seyfert 1 and 2
galaxies (see discussions in Mushotzky \etal 1995; Tanaka \etal 1995; 
Fabian \etal 1995; Yaqoob \etal 1995b).
However, if the line rest-frame energy is
really $\sim 6.7$ keV or higher then an ionized disk would account for the
large equivalent width (e.g. Matt \etal 1993). A possible
problem with this interpretation is the low accretion rate compared
to Eddington. Note that the data can accommodate the reflected continuum
from such a disk but cannot constrain it.

\subsection{Conclusions}

The X-ray continuum of M81 appears to have a nonthermal, power-law 
form and we find little evidence of any additional, soft X-ray
continuum.
Some of the X-ray properties of the AGN in M81 are similar to classical
Seyfert 1 galaxies whilst others have more in common with higher
luminosity AGN, or quasars. The properties of the broad iron line in M81
are similar to those recently found by \asca for several Seyfert 1
galaxies. The broadening mechanism, namely Doppler
and/or gravitational redshifts in matter close to the X-ray
source may also be similar in M81 and Seyfert 1s. If the matter is
in the form of a disk then the inclination angle relative to
the observer is $\sim 40$ degrees. The photon index 
($\Gamma \sim 1.9$) of the power-law
continuum is similar to that observed in quasars and is also similar
to that thought to be the {\it intrinsic} photon index 
in Seyfert 1s (e.g. Nandra \& Pounds 1994). Thus the 
origin of the hard X-ray continuum {\it may be the same} in LINERS,
Seyfert galaxies and quasars, with the caveat that in Seyfert 1s the
{\it observed} continuum has been reprocessed by matter out of
the line-of-sight and in Seyfert 2s the observed continuum is 
most affected by line-of-sight reprocessing.

AGN with X-ray luminosities of the order $\sim 10^{41} - 10^{42}
\rm \ erg \ s^{-1}$ typically vary rapidly by factors $\sim 2-3$
on a timescale of hours or less, whilst higher luminosity AGN
are more sluggish. On other hand, rapid variability is not typical
of the AGN in M81. This is puzzling, suggesting that rapid
variability only occurs for a restricted range in AGN luminosities,
covering 2-3 decades.

\section{Discussion}
\begin{figure}[htbp]
%\plotfiddle{/local/home/neutron/ptak/galaxies/lumin_ap_idl.ps}
\plotfiddle{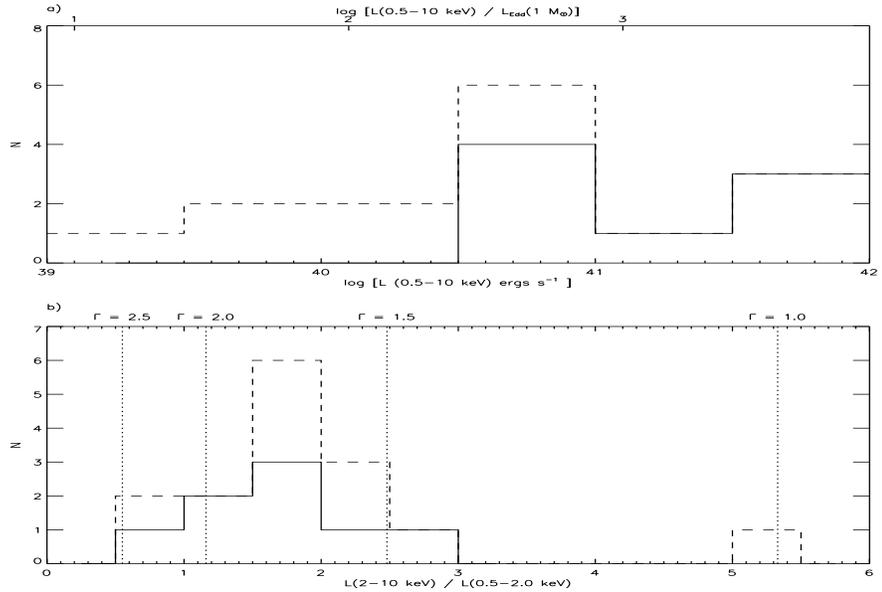}
{3.in}{0}{70}{40}{-225}{-50}
\caption{\small {\bf a)} 0.5-10 keV luminosities of the galaxy sample.  
Solid line shows 
LINERs/LLAGN only.  Top axis shows log of number of solar-mass binaries 
emitting at the Eddington luminosity required to produce the luminosities
shown. {\bf b)} As in a) but with hardness ratios.  The dotted lines show the 
expected hardness ratio from an absorbed power-laws with photon indices of
2.5, 2.0, 1.5 and 1.0
\normalsize}
\end{figure}
\begin{figure}[htbp]
%\plotfiddle{/local/home/neutron/ptak/galaxies/halpha_gamma_ap_idl.ps}
\plotfiddle{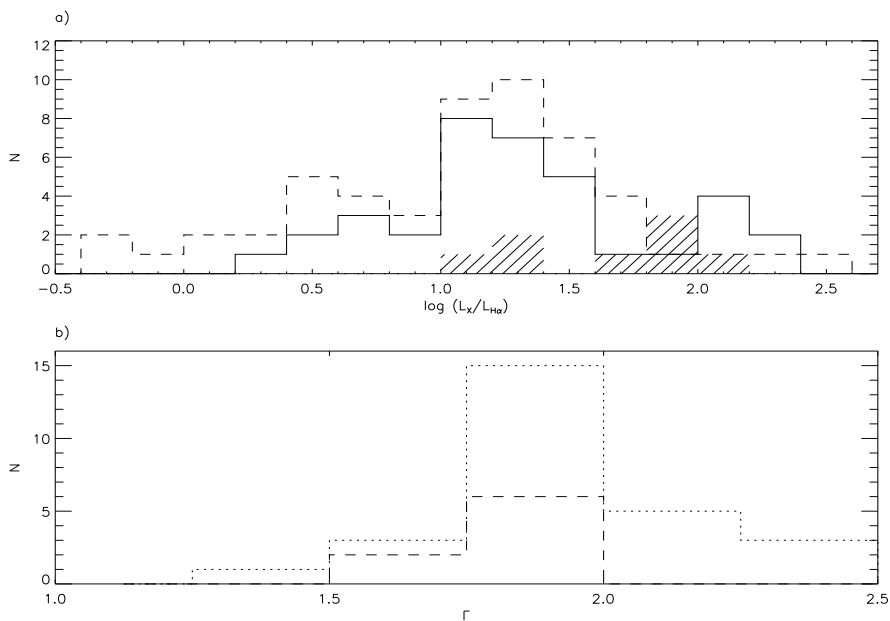}
{2.65in}{0}{70}{65}{-220}{-250}
\caption{\small {\bf a)} Distribution of $\rm L_X / L_{H\alpha}$ for AGN
observed in the 0.5-4.5 keV band (dashed line) and 2-10 keV band (solid line) 
from Elvis, Soltan, \& Keel (1984), with the filled bars showing the \asca
results in the 2-10 keV band for 8 LINERs/LLAGN.
{\bf b)} Distribution of photon indices ($\Gamma$) observed in the 
power-law
component of LINERS/LLAGNs (dashed line) and the Nandra \& Pounds (1994) 
sample of Seyfert 1 AGN (dotted line) (the Nandra \& Pounds (1994) slopes
are from fits including Compton reflection)
\normalsize}
\end{figure}

The hard X-ray emission of both starburst and LINER galaxies is
usually dominated by a compact nuclear, or near-nuclear source and there
are often one to several other off-nuclear
X-ray sources in the galaxy which together
constitute the bulk of the X-ray luminosity.
Figure 10a shows a histogram of the luminosities observed in the entire galaxy
sample, with LINERs/LLAGN (NGC 3147, NGC 3998, NGC 4579, NGC 4594, NGC 4258,
M51, NGC 3079, M81) plotted separately.  LINER/LLAGN are more
luminous than the starburst galaxies, in general.  In both cases, tens to
thousands of solar-mass X-ray binaries radiating at the Eddington limit would
be required to produce the observed flux.  Also shown (Figure 10b) is a 
histogram
of the hardness ratio (2-10 keV flux / 0.5-2.0 keV flux)
from which it can be seen that there is no clear separation between the
starbursts and the LINER/LLAGN. Most of the galaxies have a hardness 
ratio similar that expected from an unabsorbed power-law with $\Gamma 
\sim 1.5-2.5$. An exception
is NGC 3628 (traditionally classified as a starburst)
which has an exceptionally flat X-ray 
spectrum with $\Gamma \sim 1.2$ (Yaqoob \etal 1995a).

Variability has been observed in 
both LINER and starburst galaxies on timescales
of weeks to years. This implies that in both starbursts and LINERs, a 
significant amount of the emission is originating from a compact ($\ll 1$ pc)
region, most likely due to a single object such as a LLAGN or X-ray binary 
(although such a binary would have to have
a much larger mass than known X-ray binaries, typically $\sim 10-1000 \rm \
M_{\sun}$. On the other hand, rapid variability, on timescales of a
day or less, is {\it not} observed. Thus, if LLAGN are
simply low-luminosity versions of classical AGN, it is puzzling that
rapid variability is most common in objects with 2--10 keV
luminosity in the range $\sim 10^{42} - 10^{43} \ \rm erg \ s^{-1}$
but vanishes below $\sim 10^{41} \ \rm erg \ s^{-1}$.

These results strongly imply a connection between 
starburst and LINER 
activity, with the soft component most likely being produced by warm gas
with $kT \sim 0.6-0.8$ keV,
possibly from an SNR-heated ISM and  starburst-driven winds.
In some cases, the hard component of both starburst
and LINERs may also be due to starburst activity, possibly resulting from
compact supernovae (cSNR).
It is possible that some LINERs (and starbursts) may be powered by
AGN-type accretion.  To test this hypothesis we have plotted histograms
of the ratio of X-ray flux to $\rm H_{\alpha}$ flux and the photon indices 
($\Gamma$) observed in the LINERs/AGN along with the same quantities for AGN 
in Figures 11a and 11b, respectively.  There is no obvious
distinction between the two groups.   Unfortunately, our sample is too small
to statistically test if the two samples originated from the same parent
population, however, Figures 11a and 11b suggest that this is the case.

The X-ray spectra of LINERs and starbursts are not very different
to those of classical AGN. The hard power-law slopes are similar to
those of quasars and the inferred {\it intrinsic} slopes of Seyfert
galaxies. It appears that quasars and the low-luminosity spiral galaxies
are devoid of large amounts of matter residing in the nucleus, which
is responsible for reprocessing the X-ray continua of the intermediate
luminosity Seyfert galaxies (e.g. Nandra \& Pounds 1994). Even the
soft extended thermal emission which is common in many of the LINERs
and starbursts has been observed in classical AGN in which the 
hard power-law is heavily absorbed, allowing the soft component to
be detected (e.g. NGC 4151 and Mkn 3; see Figure 4).

Another puzzling result is the lack of significant Fe-K 
line emission in galaxies
like NGC 253 and M82 which are thought to be prototypical starburst
galaxies in which case the hard X-ray emission would be expected to
have a thermal origin. Yet, the lack of Fe-K line emission and the
OSSE detection of NGC 253 (Bhattacharya \etal 1994) suggest a nonthermal
origin. Fe-K line emission 
(in all cases, likely to be due
to fluorescence) is clearly detected in only three of the
galaxies: NGC 3147 (Ptak \etal 1996), NGC 4258 (Makishima \etal 1994)
and M81 (Ishisaki \etal 1996 and \S 6). In the first two cases the
Fe-K line very likely originates in obscuring matter around the nucleus,
in close analogy to the mechanism in Seyfert 2 galaxies. In the case
of M81 the characteristics of the Fe-K line are very similar to those
in Seyfert 1 galaxies and the line may indeed have the same origin;
an X-ray illuminated accretion disk. The inferred low accretion rate
of M81 may however be problematic. Much work remains to be done in 
understanding the physical implications of the X-ray emission in 
low-luminosity spiral galaxies.

\acknowledgments
We thank Rich Mushotzky and Hagai Netzer for useful discussions.  We also
would like to thank Gail Reichert for sharing her results prior to publication.
This research has made use of data obtained through the High Energy
Astrophysics Science Archive Research Center Online Service, provided by the
NASA-Goddard Space Flight Center and the NASA/IPAC Extragalactic Database
(NED) which is operated by the Jet Propulsion Laboratory, Caltech, under
contract with NASA.

\end{document}